\documentclass[useAMS,usenatbib,usegraphicx]{mn2e}
\usepackage{amstext}
\usepackage{amssymb}
\usepackage{amsmath}

\providecommand{\tabularnewline}{\\}

\def\apj{ApJ}
\def\apjl{ApJL}

\def\jcap{{ J.\ Cosmol.\ Astropart.\ Phys.\ }}

\def\mnras{MNRAS}

\def\prd{Phys.\ Rev.\ D }

\title[Nonlinear Bias of Cosmological Halo Formation]{Nonlinear Bias of Cosmological Halo Formation in the Early Universe}

\author[Ahn et al.]
{Kyungjin Ahn,$^1$\thanks{E-mail: kjahn@chosun.ac.kr} 
  Ilian T. Iliev,$^2$ 
  Paul R. Shapiro$^3$ and
  Chaichalit Srisawat$^2$ \\
  $^1$Department of Earth Sciences, Chosun
    University, Gwangju 501-759, Korea\\
  $^2$Astronomy Centre, Department of Physics and Astronomy, Pevensey II Building, University of Sussex, Falmer, Brighton BN1 9QH\\
  $^3$Department of Astronomy, University of Texas, Austin, TX
    78712-1083, USA}

\begin{document}

\maketitle

\label{firstpage}

\begin{abstract}
We present estimates of the nonlinear bias of cosmological halo formation,
spanning a wide 
range in the halo mass from $\sim10^{5}{\rm M}_{\odot}$ to
$\sim10^{12}{\rm M}_{\odot}$, 
based upon both a suite
of high-resolution cosmological N-body simulations and theoretical
predictions.
The halo bias is expressed in terms of the mean bias and stochasticity
as a function of local overdensity ($\delta$), under different
filtering scales, which is realized as the density of individual cells
in uniform grids. The sampled overdensities span a range wide enough
to provide 
the fully nonlinear bias effect on the formation of haloes. A 
strong correlation between $\delta$ and halo population overdensity
$\delta_{h}$ is found, along with sizable stochasticity.
We find that the empirical mean halo bias matches, with good accuracy,
the prediction by the peak-background split method based on the excursion
set formalism, as long as the empirical, globally-averaged halo mass function is
used. Consequently, this bias formalism is insensitive to uncertainties
caused by varying halo identification schemes, and can be applied
generically. We also find that the probability distribution function
of biased halo numbers has wider distribution than the pure Poisson
shot noise, which is attributed to the sub-cell scale halo correlation.
We explicitly calculate this correlation function and show that both
overdense and underdense regions have positive correlation, leading
to stochasticity larger than the Poisson shot noise in the range of
haloes and halo-collapse epochs we study.
\end{abstract}


\section{Introduction}
\label{sec:intro}

In the standard scenario of cosmological structure formation, cosmological
haloes are the features of the cosmic web of highest overdensity in
approximate virial equilibrium, that result from the nonlinear
amplification of initially-linear, Gaussian-random density
fluctuations by gravitational instability.
Galaxies and haloes, however, are not unbiased tracers
of the underlying density distribution, and thus understanding this
{}``bias'' effect is crucial to extract cosmological information
from the data of galaxy surveys, for example.

The idea that galaxy bias (from this point on, we will sometimes denote
dark matter haloes loosely by {}``galaxies'' to reflect the original
ideas of associating galaxies purely by high-density peaks without
resorting to hydrodynamical cooling mechanism) exists and can be calculated
from the statistics of Gaussian random initial density fields was
pioneered by \citet{Kaiser1984}. \citet{Bardeen1986} extended this
idea to take a full account of the Gaussian random density field in
a cosmological context, to understand how haloes grow out of this random
field and cluster spatially. In the meantime, \citet[PS hereafter]{1974ApJ...187..425P}
associated cosmological haloes (or galaxies) as high-density peaks
and estimated halo mass function, and this PS formalism was recounted
more rigorously by \citet{Bond1991} through their excursion set formalism
(sometimes called the extended PS formalism), where they showed that
cloud-in-cloud effect explains the fudge multiplicity factor 2 in
the PS mass function. All these ideas form the backbone of the peak-background
split scheme for calculating the galaxy bias by \citet{Cole1989},
which bears the idea that haloes (peaks) are more typically formed
in high density regions. \citet[MW hereafter]{Mo1996} calculated fully nonlinear bias combining
the peak-background split scheme with the spherical top-hat collapse
model under the extended Press-Schechter formalism, and also calculated
the useful {}``linear bias parameter'' in the linear regime. 
The peak-background split scheme may not give a perfectly accurate
prediction of N-body simulation results (e.g. MW; \citealt{Manera2010}),
which is usually attributed to the discrepancy between the PS mass
function and the N-body halo mass function at low- and high-mass ends
(e.g. \citealt{Sheth1999}, ST hereafter; \citealt{2001MNRAS.321..372J}).
This discrepancy stimulated better-fitting functional
forms (e.g. ST; \citealt{2001MNRAS.321..372J,Warren2006,Reed2007,Lukic2007,Lim2013,Watson2014}).
\citet[BL hereafter]{2004ApJ...609..474B} then developed a hybrid
scheme of combining ST mass function and the linear bias parameter
derived from the extended Press-Schechter formalism and showed that
this fitted the linear N-body halo bias better than MW prediction.

Bias
can of course have stochasticity, which was formulated theoretically
by \citet[DL henceforth]{Dekel1999}: haloes sampled inside a suite of Eulerian cells
of a given density, or count-in-cell (CIC) haloes, are expected to
deviate from purely Poisson distribution, if there is either
correlation or anti-correlation of haloes at sub-cell scales which then
result in variance of the number of haloes ($\sigma^{2}(N)$) larger or smaller than
Poissonian value, respectively
(e.g. \citealt{1993ppc..book.....P}; see also
Section~\ref{sub:Statistics}). \citet{Somerville2001} compared the
prediction by DL to N-body simulation results, and based
on the observed $\sigma^{2}(N)$ they concluded that haloes are usually correlated 
in overdense regions and anti-correlated in underdense regions (we
will however contradict this claim in Section 
\ref{sub:Result:scatter}). 
Later work found that haloes usually show variance larger than the
Poissonian value (e.g. \citealt{Neyrinck2014} find that haloes of mass
$10^{10-11}\,M_{\odot}$  show this ``super-Poissonian'' distribution
under $2/h\,{\rm Mpc}$ cells), which are well fitted by the functional
distributions suggested by \citet{Saslaw1984} and
\citet{Sheth1995}. 

A useful application of the nonlinear halo bias prescription is to create mock halo
catalogues in a large scale for either cosmology or
astrophysics. While mock galaxy catalogues can be created by
schemes based on quasi-linear perturbation theory, such as PINOCCHIO \citep{Monaco2002,Monaco2013}
and PTHALOES \citep{Scoccimarro2002,Manera2013}, they are usually limited to the scales under which
density perturbation remains quasi-linear at most. This limitation can
be overcome by nonlinear halo bias schemes, as in
\citet{Kitaura2014} who prove the concept by generating halo
catalogues which are statistically consistent with N-body halo
catalogues, suited for probing the baryon acoustic oscillation (BAO)
feature by surveys such as the Baryon Oscillation Spectroscopic Survey
(BOSS). 
We intend to achieve a similar goal in the long run, but with
a bias scheme that is fully nonlinear  and is applicable regardless of
the halo mass, the filtering scale and the redshift. Because we will
calculate the bias parameter theoretically, our scheme will mitigate
the need to find an empirical fitting formula as done in
e.g. \citet{Kitaura2014}.

A similar formalism can also be applied to astrophysical problems. 
Understanding the halo bias is crucial e.g. in the study of cosmic
reionization, due to the very large dynamic gap between the very small
galaxies believed to be the main drivers of reionization \citep[see e.g.][for a review]{2005SSRv..116..625C}
and the large characteristic scales of the reionization patchiness
\citep{Friedrich2011,Iliev2014}. \citet{2004ApJ...609..474B} used a hybrid
halo bias scheme  to study the fluctuation
of the 21cm background from the fluctuating halo distribution during
the epoch of reionization (EoR). 
Fast semi-numerical simulators of reionization
\citep{Zahn2007,Santos2008,Alvarez2009,Mesinger2011}, whose basis was 
formulated by \citet{2004ApJ...613....1F} and \citet{2005MNRAS.363.1031F} to
replace the time-consuming ray-tracing by a faster excursion set formalism, make
use of a similar formalism to seed haloes in a coarse-grained density
field.

We have indeed applied this formalism to a simulation of cosmic
reionization, by which we could span the full dynamic range of
halos hosting radiation sources.
 Cosmic reionization is 
believed to occur very inhomogeneously with large H II regions,
whose sizes show a wide distribution peaked at $\sim 20$ comoving
Mpc before completion if roughly put. Therefore it is necessary to use a
large box in order to simulate the reionization process in a statistically
reliable way. This requirement, however, limits the ability of the
simulation to resolve 
``minihaloes'' which are believed to host Population III stars, and allows
the simulation to only resolve  the more massive kind, or
``atomic-cooling halos''.
Indeed most reionization simulations in large boxes used
to implement atomic-cooling halos only, while this may
underestimate the photon budget in the early stage of reionization.
In a large-scale (box size of $114/h\,{\rm Mpc}$
comoving) simulation of cosmic reionization (with ray-tracing method), \citet{Ahn2012} used
the conditional halo bias found in Section \ref{sub:Minihaloes}
of this paper to include minihaloes, which could not otherwise have
been realized due to numerical resolution. This way, they could span
the full dynamic range of haloes -- both minihaloes and atomic-cooling
haloes -- responsible for emitting hydrogen-ionizing and ${\rm
  H}_2$-dissociation radiation, and
observed that the reionization process is extended further in
time to comply better with several observational constraints.

On much larger scales (box size of $425/h\,{\rm Mpc}$ comoving) the
same technique was used to perform the largest-volume, ray-tracing simulations
of cosmic reionization  to date, presented in \citet{Iliev2014} and
further explored in \citet{Datta2012},
\citet{Park2013} and \citet{Shapiro2013}. This used the results in
Section~\ref{sub:Atomhalo} to include the 
unresolved low-mass atomic cooling haloes ($M=10^{8}-10^{9}\, M_{\odot}$).
Another prospective application is in exploring the effects of primordial
non-Gaussianity on halo bias, which is an active area of research
(e.g. \citealt{Dalal2008}; \citealt{Adshead2012}; \citealt{DAloisio2013}),
and which also leads to ionization bias (e.g. \citealt{Joudaki2011,DAloisio2013})
detectable by 21 cm observations (e.g. \citealt{Mao2013}).


In this paper, we examine and compare the nonlinear halo bias from
both our suite of cosmological N-body simulations suited for the study
of haloes responsible for EoR and a semi-analytical, fully nonlinear
peak-background split scheme. This theoretical scheme is a hybrid
scheme similar to the one by BL, but also differ as we combine
the empirical (mean) halo mass function to the bias factor and  extend
it to the fully nonlinear regime in a non-perturbative way. Through
this, we investigate whether the bias factor can be purely based upon
the excursion set formalism and separated cleanly from the mass function,
which bears uncertainty due to its strong dependence on specific halo-identification
schemes. We also study the stochasticity of halo bias from these simulations
and examine whether they are purely Poissonian or not, which has been
investigated recently to conclude that haloes in some mass range indeed
have super-Poissonian distribution
\citep{Baldauf2013,Neyrinck2014}. Toward this, 
we calculate the 2-point halo correlation function and quantify its
contribution to stochasticity in addition to the Poisson noise. While
our paper is focused on the range of haloes responsible for cosmic
reionization at $z\gtrsim6$, and therefore it can be used readily
in the study of EoR, our formalism should be applicable in more
generic cases.

This paper is organized as follows. In Section~\ref{sec:nbody}, we
briefly describe our N-body simulation. In Section~\ref{sec:Theory},
we describe the theoretical scheme for the nonlinear halo bias, which
combines the peak-background split scheme (Sections
\ref{sub:bias-linear} and \ref{sub:bias-NL}) with the empirical N-body
halo mass function (Section~\ref{sub:bias_and_hybrid_mass_function}),
and also describe the stochasticity and various quantities related
(\ref{sub:Statistics}). We then describe our results
in Section~\ref{sec:Result}, first on the mean halo mass function
(Section~\ref{sub:Result:Halo-Mass-Function}), then on the mean bias
(Section~\ref{sub:Result:mean}) and on the stochasticity
(Section~\ref{sub:Result:scatter}). We further investigate the
validity of the usual linear bias approximation in
Section~\ref{sub:Validity-of-Linear}. We conclude our paper in
Section~\ref{sec:conclusion}, together with a schematic layout of our
bias prescription toward generating mock halo catalogues.





\section{Simulations}
\label{sec:nbody}

The data used in this work is based on a suite of large simulations,
most of which were previously presented in \citet{Watson2014}. They
were performed using the CubeP$^{3}$M code, a high-performance, publicly
available, cosmological N-body code based on particle-particle-particle-mesh
(P$^{3}$M) scheme (for detailed code
description and tests see \citealt{Harnois-Deraps2013}). 
For memory efficiency and speed the code uses two-level grid for computing
the long-range gravity forces using a particle-mesh method and adds
the local direct particle-particle forces at small scales. CubeP$^{3}$M
is a massively parallel, hybrid (using MPI and OpenMP) code, scaling
well up to tens of thousands of computing cores. It has been extensively
tested and run on a wide variety of parallel platforms.

Our complete simulation suite, listed in Table~\ref{summary_N-body_table},
includes volumes between $6.3/h\,$Mpc and $114/h\,$Mpc
per side and between $1728^{3}$ and $5488^{3}$ particles, thereby
covering a large dynamic range, with particle masses ranging from
$5.2\times10^{3}\, M_{\odot}$ to $5.5\times10^{6}\, M_{\odot}$ and force smoothing
lengths between $182\,$pc and $1.86/h\,$kpc. The smaller-volume,
high-resolution simulations with boxes up to $20/h\,$Mpc per
side resolve (with 20 particles or more) dark matter haloes with mass
$10^{5}\, M_{\odot}$ and above, the expected hosts of the First Stars.
In contrast, the larger volume, $114/h\,$Mpc
only resolves haloes with mass $10^{8}\, M_{\odot}$ (with 20
particles) and larger, but samples the statistics
of rare haloes much better due to its  larger volume.

We locate the collapsed haloes at runtime, using the CPMSO spherical
overdensity method \citep{Harnois-Deraps2013,Watson2014} with overdensity
with respect to the mean of 178, suitable for the high redshifts
considered here. This is done by first interpolating
the particles onto a fine grid (with number of cells per dimension
twice the number of particles) using the cloud-in-cell (CIC) approximation.
Local density peaks (with density at least 100 times the average)
are located and spherical shells are expanded around each peak until
the threshold overdensity is crossed. The resulting object is then
marked as a halo (objects with less than 20 particles are discarded
as they cannot be reliably identified). The halo centre position is
calculated more precisely by quadratic interpolation within the cell
and the particles within the halo virial radius are identified and
then the halo properties, e.g. mass, velocity dispersion, centre-of-mass,
angular momentum, radius, etc. are calculated and saved in the halo
catalogue.

\begin{table*}
 \caption{N-body simulation parameters. Background cosmology is based on the
WMAP 5-year results: $\Omega_{m}=0.27$,
$\Omega_{\Lambda}=0.73$, $Omega_{b}=0.044$, $h=0.7$, $\sigma_{8}=0.8$, and $n_{s}=0.96$.}


\label{summary_N-body_table}

\centering{}\begin{tabular}{@{}l@{}llllll}
\hline 
simulation$\,$  & box size  & $N_{\rm particle}$  & mesh  & spatial resolution  & $m_{\rm particle}$  & $M_{\rm halo,min}$ \tabularnewline
\hline
S1  & 6.3$\, h^{-1}$Mpc  & $1728^{3}$  & $3456^{3}$  & $182\,{h^{-1}}$pc  & $5.19\times10^{3}\, M_{\odot}$  & $1.04\times10^{5}\, M_{\odot}$ \tabularnewline
M1  & 20$\, h^{-1}$Mpc  & $5488^{3}$  & $10976^{3}$  & $182\,{h^{-1}}$pc  & $5.19\times10^{3}\, M_{\odot}$  & $1.04\times10^{5}\, M_{\odot}$ \tabularnewline
B1  & 114$\, h^{-1}$Mpc  & $3072^{3}$  & $6144^{3}$  & $1.86\,{h^{-1}}$kpc  & $5.47\times10^{6}\, M_{\odot}$  & $1.09\times10^{8}\, M_{\odot}$ \tabularnewline
\hline
\end{tabular}
\end{table*}

\section{Theory}
\label{sec:Theory}

Formation of cosmological haloes is strongly correlated with their
larger-scale density environment. The excursion set formalism (\citealt{Bond1991})
gives a quantitative description of this biased halo formation in
terms of the conditional halo mass function $dn/dM\,(M;\,\delta)$,
where $\delta\equiv(\rho-\bar{\rho})/\bar{\rho}$ is the overdensity
of the local environment. This description is called the peak-background
split, where haloes are considered as the high-density {}``peaks''
that are placed on large-scale density {}``background''. In the
linear regime where $\delta\ll1$, this yields the linear bias parameter
which has been used extensively in cosmology (MW).

In this Section we introduce a formalism which is intended to describe
the local nonlinear bias in a non-perturbative way, based mostly on
the formalism by MW and the idea of BL. Therefore, we revisit previous
theoretical work, and at the same time describe modifications we made
in this Section. We will then compare the prediction from this formalism
to the N-body data results in Section \ref{sec:Result}.  We will
occasionally add subscript ``$L$'' to Lagrangian quantities, when
otherwise these may be confused with Eulerian ones.

\subsection{Biased halo mass function in Lagrangian volume }
\label{sub:bias-linear}

It is shown in the excursion set formalism that distribution of linear
overdensity $\delta$ in the initially Gaussian-random matter density
field $\rho_{L}$
filtered with a {}``sharp $k$-space filter'',

\begin{eqnarray}
\rho_{L}({\bf r},\, R_{f,L})&=&
\int d^{3}r'W_{K}({\bf r}-{\bf r'};\,
R_{f,L})\rho_{L}({\bf r'},\,0),\nonumber \\
\rho_{L}(k,\, R_{f,L})&=&\tilde{W_{K}}(k;\,R_{f,L})\rho_{L}({\bf k},\,0),
\label{eq:filtered_density}
\end{eqnarray}
 where the window function $W_{K}(r;\, R_{f,L})$ is the Fourier transform
of sharp $k$-space filter $\tilde{W}_{K}(k;\, R_{f,L})\equiv\Theta(1-kR_{f,L})$,
still follows Gaussian distribution (\citealt{Bond1991}). This is obviously
true even in the density field linearly extrapolated to the observing
redshift with the linear growing factor. 
This way, one can use the linearly extrapolated density field and the appropriate
halo collapse criterion to predict halo population at any filter scale
and redshift.

In this formalism the unconditional, globally-averaged differential halo
number density (mass function) is given by the Press-Schechter formula
(\citealt{1974ApJ...187..425P}) 
\begin{eqnarray}
\left(\frac{dn}{dM}\right)_{{\rm
    PS}}(M)&=&\left(\frac{dn}{dM}\right)_{{\rm
    PS}}(\sigma_{M,L}^{2};\delta_{c}) \nonumber \\
&=&-\frac{1}{\sqrt{2\pi}}\frac{d\sigma_{M,L}^{2}}{dM}\frac{\bar{\rho}_{0}}{M}\frac{\nu}{\sigma_{M,L}^{2}}\exp\left[-\frac{\nu^{2}}{2}\right],
\label{eq:ps}
\end{eqnarray}
 where $\sigma_{M,L}^{2}$ is the variance of Gaussian distribution
of the density field (linearly extrapolated to the present) filtered
in real space in spheres with radius $R_{f,L}$, $\bar{\rho}_{0}$ is
the present matter density, $\nu\equiv\delta_{c}/(D(z)\sigma_{M,L})$
(with the linear growth factor $D(z)$ in $\Lambda$CDM universe)
is the ratio of critical overdensity $\delta_{c}=1.686$\footnote{We
  neglect the very weak redshift dependence of $\delta_{c}$ in
  $\Lambda$CDM in our study, while for $z\lesssim 4$ one should implement
  its redshift dependence.} to
$\sigma_{M,L}(z)=D(z)\sigma_{M,L}$,
and $R_{f,L}$ is the length scale usually associated%
\footnote{Rigorously speaking, we cannot associate such a well-defined mass
$M$ with $R_{f,L}$ in the case of sharp-$k$ filtering (e.g. \citealt{Bond1991}).
However, we adopt this definition for simplicity.%
} with the halo mass $M$ by 
\begin{equation}
M=M(R_{f,L})=\bar{\rho}_{0}\frac{4\pi}{3}R_{f,L}^{3}.
\label{eq:mass-R}
\end{equation}
Equation (\ref{eq:ps}) is unconditional in a sense that this represents
the average halo distribution in the universe.

The excursion set formalism also predicts halo population inside a
region with given mean overdensity and size. Sharp $k$-space filtering
allows one to write this in terms of conditional probability analytically,
because wavemodes at different filter scales are linearly independent.
The barrier crossing and variance under given density environment,
which will be denoted by a {}``cell'', is measured from the new
origin $\delta_{{\rm lin}}$ (throughout this paper, unless specified
differently, we denote the full, nonlinear overdensity of a cell by
$\delta$ for simplicity) and $\sigma_{{\rm cell},L}$, which are
linearly extrapolated density of the cell and variance corresponding
to the Lagrangian cell size $R_{{\rm cell},L}$, respectively. When an Eulerian cell has
a comoving volume $V_{{\rm cell}}$ and nonlinear overdensity
$\delta$ at some redshift, $R_{{\rm cell},L}$ can be obtained from
\begin{equation}
M_{\rm cell} =\bar{\rho}_{0}V_{{\rm cell}}\left(1+\delta\right)
=\bar{\rho}_{0}\frac{4\pi}{3}R_{{\rm cell},L}^{3}.
\label{eq:mass-Rcell}\end{equation}
According to the well-known excursion-set formalism (\citealt{Bond1991}),
the differential halo number density (halo mass function) inside a Lagrangian
region with $\delta_{{\rm lin}}$ (linearly extrapolated to redshift
$z$) and $R_{{\rm cell},L}$ is then given 
by a conditional mass function 
\begin{eqnarray}
\left(\frac{dn}{dM}\right)_{{\rm PS},\, b}^{{\rm L}}(M|\delta_{{\rm lin}}) & \equiv & \left(\frac{dn}{dM}\right)_{{\rm PS}}(\sigma_{M,L}^{2};\,\delta_{c}|\sigma_{{\rm cell},L}^{2};\,\delta_{{\rm lin}})\nonumber \\
 & = & \left(\frac{dn}{dM}\right)_{{\rm PS}}(\sigma_{M,L}^{2}-\sigma_{{\rm cell},L}^{2};\,\delta_{c}-\delta_{{\rm lin}})\nonumber \\
 & = &
-\frac{1}{\sqrt{2\pi}}\frac{d\sigma_{M,L}^{2}}{dM}\frac{\bar{\rho}_{0}}{M}
\frac{\left(\delta_{c}-\delta_{{\rm lin}}\right)/D(z)}{\left(\sigma_{M,L}^{2}-\sigma_{{\rm cell},L}^{2}\right)^{3/2}} \times \nonumber \\
&&\exp\left[-\frac{\left(\delta_{c}-\delta_{{\rm lin}}\right)^{2}}{2D^{2}(z)\left(\sigma_{M,L}^{2}-\sigma_{{\rm cell},L}^{2}\right)}\right]
\label{eq:ps-bias}
\end{eqnarray}
which takes the same form as equation (\ref{eq:ps}) but with $\sigma_{M,L}^{2}$
and $\delta_{c}$ replaced by $\sigma_{M,L}^{2}-\sigma_{{\rm cell},L}^{2}$
and $\delta_{c}-\delta_{{\rm lin}}$, respectively. 
Here, $\sigma_{{\rm cell},L} \equiv \sigma_{M_{\rm cell},L}$.
This defines the Lagrangian
over-abundance of haloes of mass M, 
\begin{equation}
\delta_{{\rm h}}^{{\rm L}}(M|\delta_{{\rm
    lin}})\equiv\left(\frac{dn}{dM}\right)_{{\rm PS},\, b}^{{\rm 
    L}}(M|\delta_{{\rm lin}})\Bigg/\left(\frac{dn}{dM}\right)_{{\rm
    PS}}(M)-1. 
\label{eq:deltaH_lagrangian}
\end{equation}

Note that two important factors should be considered in order to generalize
equation (\ref{eq:ps-bias}). First, in the nonlinear regime where
$\delta_{{\rm lin}}\sim 1$, one should match the
nonlinear $\delta$ to the linear $\delta_{{\rm lin}}$
to use equation (\ref{eq:ps-bias}), because this is based on the
linear theory. Second,$\left(dn/dM\right)_{{\rm PS},\, b}^{{\rm L}}$
and $\delta_{{\rm h}}^{{\rm L}}$ should be converted into the corresponding
Eulerian mass function and Eulerian halo over-abundance, respectively,
because Eulerian quantities are of much more practical use than Lagrangian
quantities. This conversion will be described in Section \ref{sub:bias-NL}.

\subsection{Nonlinear background and biased haloes mass function in Eulerian volume}
\label{sub:bias-NL}

The quantities $\left(dn/dM\right)_{{\rm PS},\, b}^{{\rm L}}$
and $\delta_{{\rm h}}^{{\rm L}}$ in equations (\ref{eq:ps-bias})
and (\ref{eq:deltaH_lagrangian}) are derived assuming that density
grows linearly with the linear growth factor and are defined in the
Lagrangian volume. In reality, growth of density perturbations is nonlinear
in general, and this also yields large difference between the Lagrangian
and Eulerian volumes.

Therefore, we first need to map nonlinear overdensity $\delta$ to
linear overdensity $\delta_{{\rm lin}}$. We use the mapping scheme
based on the tophat collapse model, which has also been used by MW, where
$\delta$, which is nonlinear in general, is linked to $\delta_{{\rm lin}}$
in a parametric form of $\theta$ as follows: 
\begin{equation}
\delta=\left(\frac{10\delta_{{\rm
      lin}}}{3(1-\cos\theta)}\right)^{3}-1,\,\,\,\delta_{{\rm
    lin}}=\frac{3\times6^{2/3}}{20}\left(\theta-\sin\theta\right)^{2/3},
\label{eq:lin-NL-pos}
\end{equation}
 if $\delta>0$. Similarly, if $\delta<0$, 
\begin{equation}
\delta=\left(\frac{10\delta_{{\rm
      lin}}}{3(\cosh\theta-1)}\right)^{3}-1,\,\,\,\delta_{{\rm
    lin}}=\frac{3\times6^{2/3}}{20}\left(\sinh\theta-\theta\right)^{2/3}.
\label{eq:lin-NL-neg}
\end{equation}
 Note that $\delta$ increases monotonically as $\delta_{{\rm lin}}$
increases, such that there exists one-to-one mapping.

We also need to consider the change of Lagrangian volume by multiplying
the ratio of Lagrangian volume to the Eulerian volume to obtain the
correct Eulerian number density, which yields the final form: \begin{equation}
\left(\frac{dn}{dM}\right)_{{\rm PS},\, b}=\left(\frac{dn}{dM}\right)_{{\rm PS},\, b}^{{\rm L}}(1+\delta).\label{eq:NL_lin_dndM}\end{equation}
 By taking further approximation that $\delta_{{\rm c}}\gg\delta$
and $\sigma_{M}\gg\sigma_{{\rm cell}}$ MW find a useful linear
relation between $\delta_{{\rm h}}$ and $\delta_{{\rm cell}}$. This
approximation implies that total mass contained in haloes inside a
cell is much smaller than the total mass of the cell. However, this
approximation is not always valid at high resolution because some
cells in our density field, depending on the choice of the cell-size,
may achieve very high overdensity $\delta_{{\rm cell}}$ such that
$\delta_{c}\gtrsim\delta$ and $\sigma_{M}\gtrsim\sigma_{{\rm cell}}$.
Therefore, we just use equation~(\ref{eq:NL_lin_dndM}) in its general
form, which allows for nonlinear relation between $\delta_{{\rm h}}$
and $\delta_{{\rm cell}}$.

\subsection{Nonlinear bias and hybrid conditional mass function}

\label{sub:bias_and_hybrid_mass_function}

Before proceeding, let us define the mean conditional bias function
$b(\delta)$ (MW; DL): \begin{equation}
b(\delta)\equiv\frac{\left\langle \delta_{{\rm h}}(M|\delta)\right\rangle _{\delta_{{\rm h}}|\delta}}{\delta},\label{eq:mean_bias_def}\end{equation}
 where $\delta_{{\rm h}}(M|\delta)$ is the conditional, Eulerian
halo over-abundance, and the seemingly repetitive definition of the
average is to clarify the fact that the average is taken \emph{only}
over the cells with the given $\delta$, following the notation from
equations (3) and (4) of DL, which
is different from the average over all cells regardless of $\delta$,
or $\left\langle \,\,\, \right\rangle $. This average takes the following
integral form for any conditional function of $\delta_{{\rm h}}$
under a given $\delta$, $f(\delta_{{\rm h}})|\delta$: \begin{equation}
[f(\delta_{{\rm h}}|\delta)]\equiv\left\langle f(\delta_{{\rm h}}|\delta)\right\rangle _{\delta_{{\rm h}}|\delta}\equiv\int d\delta_{{\rm h}}P(\delta_{{\rm h}}|\delta)f(\delta_{{\rm h}}),\label{eq:bias_DL}\end{equation}
 where $P(\delta_{{\rm h}}|\delta)$ is the conditional probability
for a cell with $\delta$ to have $\delta_{{\rm h}}$ as the halo
over-abundance inside it (DL), and only those cells with given $\delta$
are included in the integration. To distinguish the \emph{conditional}
averaging from the normal averaging $\left\langle f\right\rangle $,
we denote the former by a simple notation, $[f]$, in which the dependence
on $\delta$ is assumed implicitly. Equation~(\ref{eq:bias_DL})
is equivalent to equation~(5) in DL.

Equations (\ref{eq:ps}), (\ref{eq:ps-bias}) and (\ref{eq:NL_lin_dndM})
naturally determine by how much the local halo mass function is modified.
The Eulerian over-abundance of haloes is then given by 
\begin{eqnarray}
&&\left[\delta_{{\rm h}}(M|\delta)\right]=\frac{\left(\frac{dn}{dM}\right)_{{\rm PS},\, b}(M|\delta)}{\left(\frac{dn}{dM}\right)_{{\rm PS}}(M)}-1 \nonumber \\
&&=\frac{\left(\frac{dn}{dM}\right)_{{\rm PS}}(\sigma_{M,L}^{2};\,\delta_{c}|\sigma_{{\rm cell},L}^{2};\,\delta_{lin})}{\left(\frac{dn}{dM}\right)_{{\rm PS}}(\sigma_{M,L}^{2};\delta_{c})}(1+\delta)-1,
\label{eq:boost}
\end{eqnarray}
which is equivalent to equation (19) of MW. The bias function $b$
becomes independent of $\delta$ in the linear regime where $\delta_{{\rm c}}\gg\left|\delta\right|\simeq\left|\delta_{0}\right|$
and $\sigma_{M,L}^{2}\gg\sigma_{{\rm cell},L}^{2}$, and is given
as a function of $\nu$ alone, at any given $z$: \begin{equation}
b_{{\rm lin}}(\delta)=1+\frac{\nu^{2}-1}{\delta_{c}(z)}\label{eq:bias_linear}\end{equation}
 (MW). $b_{{\rm lin}}$ is referred to as the linear bias parameter. 
We will test the applicability of this approximation in
\S~\ref{sub:Result:mean} and \ref{sub:Validity-of-Linear}.

The relation between $\delta_{{\rm h}}$ and $\delta$ is generally
nonlinear, and therefore equation (\ref{eq:bias_linear}) is of limited
use for our purposes. Even in the linear regime where $\left|\delta\right|\ll1$,
using equation (\ref{eq:bias_linear}) may be problematic because
the other condition $\sigma_{M}^{2}\gg\sigma_{{\rm cell}}^{2}$
is not valid in general and then the exponential term in equation~(\ref{eq:ps-bias})
cannot be approximated further. For example, for minihaloes of $M\ge10^{5}\, M_{\odot}$,
we have $\sigma_{M}^{2}\le70.6$, while cells we study here
have masses (when $\delta=0$) as low as
$3.5\times10^{8}\, M_{\odot}$ (in both $6.3\,h^{-1}\,{\rm Mpc}$ and
$20\,h^{-1}\,{\rm Mpc}$ boxes), 
which corresponds to $\sigma_{{\rm cell}}^{2}=24.0$.

One may naively expect that $\left(dn/dM\right)_{{\rm PS},\, b}$
gives the correct analytical estimate for the biased halo mass function.
However, it is well known that the unconditional PS mass function,
$\left(dn/dM\right)_{{\rm PS}}$, is a poor fit to the empirical
halo mass function derived from N-body simulations, in general, depending
on the range of mass -- especially so for rare haloes -- and redshift
(e.g. \citealt{2001MNRAS.321..372J}). 
It is thus reasonable to expect that $\left(dn/dM\right)_{{\rm PS},\, b}$
will also become a poor fit to the biased N-body halo mass function.

We therefore adopt a hybrid approach, first introduced by \citet{2004ApJ...609..474B},
to predict the conditional mass function (or bias) by combining $\delta_{{\rm h}}(\delta)$
(or equivalently $b(\delta)$) as in equation (\ref{eq:boost}), derived
from the excursion set formalism, with the unconditional mass function
$dn/dM$, which we choose independently. This approach is
somewhat advantageous over \citet{2002MNRAS.329...61S} and PS, for
example, because $\delta_{{\rm h}}(\delta)$ or $b(\delta)$ is almost
independent of how haloes are identified (MW) and thus the unconditional
mass function can be found empirically for any arbitrarily identified
N-body haloes. We can then expect that when such an empirical mass
function $dn/dM$ is combined with equation (\ref{eq:boost}),
the resulting mass function may be a better fit to the actual biased
halo mass function $\left(dn/dM\right)_{b}$.

In contrast to \citet{2004ApJ...609..474B}, who choose the well-known
PS and Sheth-Tormen (ST) mass functions, we choose three mass functions:
PS, ST and the empirical fit to our N-body data. The reason for using
the empirical (unconditional) mass function is because (1) both PS
and ST mass functions are known to be poor-fits to very rare haloes
(see discussion in \citealt{Watson2014} and references therein) and
for the redshift and halo mass range of interest here all haloes are
rare and (2) we want a prescription which is independent of the systematic
uncertainties of the unconditional mass function due to the varying
halo-identification schemes. The conditional PS bias trivially reduces
to $\left(dn/dM\right)_{{\rm PS},\, b}$, while in the other
two cases, the unconditional ST $\left(dn/dM\right)_{{\rm ST}}$
and the empirical fit $\left(dn/dM\right)_{{\rm N-body}}$
are both simply multiplied by $1+\delta_{{\rm h}}$ to produce 
\begin{eqnarray}
\left(\frac{dn}{dM}\right)_{{\rm ST},\, b}&=&\left\{ 1+\delta_{{\rm
    h}}(\delta)\right\} \left(\frac{dn}{dM}\right)_{{\rm ST}} \nonumber \\
&=&\left\{
1+b(\delta)\delta\right\} \left(\frac{dn}{dM}\right)_{{\rm ST}}
\label{eq:ST-bias}
\end{eqnarray}
and 
\begin{eqnarray}
\left(\frac{dn}{dM}\right)_{{\rm N-body},\, b}&=&\left\{ 1+\delta_{{\rm
    h}}(\delta)\right\} \left(\frac{dn}{dM}\right)_{{\rm
    N-body}} \nonumber \\
&=&\left\{ 1+b(\delta)\delta\right\}
\left(\frac{dn}{dM}\right)_{{\rm N-body}},
\label{eq:Nbody-bias}
\end{eqnarray}
where $\delta_{{\rm h}}(\delta)$ is given by equation (\ref{eq:boost}).
It is important to note that even when $\delta=0$, $(1+\delta_{{\rm h}})\neq1$
in general. In order to illustrate this, let us consider the limiting
case of very rare haloes such that $\nu\gg1$. Such haloes will most
likely form at very high-density regions -- or more explicitly, high-density
cells with some fixed Eulerian volume -- with $\delta\gg0$. In this
case, $(1+\delta_{{\rm h}})\to0$ or $b(\delta)\delta\to-1$ as $\delta\to0$,
and thus a simple linear relation $\delta_{h}\propto \delta$, 
which yields $(1+\delta_{{\rm h}})\to1$
as $\delta\to0$, inevitably fails in estimating the bias correctly
even in the linear regime. More detailed discussion of this aspect is in
Section~\ref{sub:Validity-of-Linear}.

Finally, the fraction of halo-mass to cell-mass, or the collapsed
fraction, is given by 
\begin{eqnarray}
f_{{\rm c},\,{\rm b}}(M_{{\rm min}},\, M_{{\rm max}})&\equiv& f_{{\rm
    c}}(M_{{\rm min}},\, M_{{\rm max}}|\sigma_{{\rm
    cell}}^{2};\,\delta) \nonumber \\
&=&\frac{\int_{M_{{\rm min}}}^{M_{{\rm
        max}}}\left(\frac{dn}{dM}\right)_{b}MdM}{\rho_{0}(1+\delta)}
\nonumber \\
&=&\frac{\int_{M_{{\rm
        min}}}^{M_{{\rm
        max}}}\left(\frac{dn}{dM}\right)_{b}^{L}MdM}{\rho_{0}},
\label{eq:fcoll_bias}
\end{eqnarray}
 which is naturally expressed in Lagrangian quantities, because both
masses inhabit the same Lagrangian region. Here once again, $\left(dn/dM\right)_{b}$
can be based on either the PS mass function, the ST mass function
or the empirical fit to simulations.

\subsection{Expected stochasticity and renormalization}

\label{sub:Statistics}

We have so far described the mean conditional mass function. In reality,
the observed correlation should exhibit stochasticity as well, because
structure forms out of a random density field. In addition, when haloes
of our interest are rare, not all the cells with given $\delta$ will
contain such haloes, giving rise to Poisson fluctuations. However, we
will soon see that the stochasticity should differ from pure
Poissonian distribution. 
Here we consider only the local stochasticity and postpone the analysis
of multi-point correlation and corresponding statistics to a future
paper.

Because the conditional mass function has a stochastic element, the
total number of haloes inside cells with given overdensity $\delta$
and Eulerian volume $V_{{\rm cell}}$ would show a scatter around the
mean value. For the total number of haloes in a mass bin $M=[M_{\rm min},\,M_{\rm max}]$, 
\begin{equation}
N(M_{{\rm min}},\, M_{{\rm max}}|\delta,\, V_{{\rm cell}})\equiv
V_{{\rm cell}}\int_{M_{{\rm min}}}^{M_{{\rm
      max}}}dM\left(\frac{dn}{dM}\right)_{{\rm o},\,{\rm cell}},
\label{eq:Ntot_biased}
\end{equation}
over different cells with the same $V_{{\rm cell}}$ and $\delta$
and where $\left(dn/dM\right)_{{\rm o},\,{\rm cell}}$ is
the observed halo mass function inside each cell, one would naively expect that the probability
distribution function (PDF) of $N$ will obey the Poisson statistics:
\begin{equation}
P_{{\rm cell}}(N)\equiv P(N|\delta,\, V_{{\rm cell}})\to\frac{e^{-\left[N\right]}\left[N\right]^{N}}{N!},\label{eq:pdf_N}\end{equation}
where the average is again taken only over the cells with given $\delta$
such that $\left[N\right]=\left\langle N\right\rangle _{\delta_{{\rm h}}|\delta}=\left\langle N(M_{{\rm min}},\, M_{{\rm max}}|\delta,\, V_{{\rm cell}})\right\rangle _{\delta_{{\rm h}}|\delta}$.
If so, both the conditional mean and conditional variance of $N$
would become identical to $\left[N\right]$. However, if correlation
of haloes at sub-cell length scale exists, there occurs an additional
variance -- either positive or negative -- in $N$ (\citealt{1993ppc..book.....P};
DL): 
\begin{equation}
\Delta_{{\rm scc}}(\delta)=\left(\frac{\left[N\right]}{V_{{\rm
      cell}}}\right)^{2}\int^{V_{{\rm
      cell}}}dV_{1}dV_{2}\,\overline{\xi_{12}}(\delta),
\label{eq:scc}
\end{equation}
where {}``scc'' denotes sub-cell correlation such that the integration
is taken inside a cell and the \emph{conditional} sub-cell 2-point
correlation function $\overline{\xi_{12}}(\delta)$ is defined by
\begin{equation}
\left[N_{1}N_{2}\right]=\left(\frac{\left[N\right]}{V_{{\rm cell}}}\right)^{2}dV_{1}dV_{2}\,\left\{ 1+\overline{\xi_{12}}(\delta)\right\} ,\label{eq:con_correl_definition}\end{equation}
where 1 and 2 denote two different sub-cell positions inside the same
cell and $N_{1}$ and $N_{2}$ are number of haloes in each sub-cell.
$\overline{\xi_{12}}(\delta)$ should not be confused with the global sub-cell
correlation function $\xi_{12}$, defined by\begin{equation}
\left\langle N_{1}N_{2}\right\rangle =\left(\frac{\left\langle N\right\rangle }{V_{{\rm cell}}}\right)^{2}dV_{1}dV_{2}\,\left\{ 1+\xi_{12}\right\} .\label{eq:normal_correl_definition}\end{equation}
Note that equation (\ref{eq:scc}) and (\ref{eq:con_correl_definition})
are restricted only to cells with given $\delta$, which are direct
applications of equations (7.66) and (7.63) in \citet{1993ppc..book.....P},
respectively. While these equations were originally intended for unconditional
quantities in \citet{1993ppc..book.....P}, applying these to conditional
quantities is trivially achieved by replacing the global average $\left\langle \,\,\,\right\rangle $
with the conditional average $\left[\,\,\,\right]$. This is easily
justified by the fact that when there is no sub-cell correlation in
those cells with $\delta$, or when
$\overline{\xi_{12}}(\delta)=0$, the identity
$[N_{1}N_{2}]=[N_{1}][N_{2}]=[N]dV_{1}/V_{\rm cell}\,[N]dV_{2}/V_{\rm
  cell}$ is satisfied by equation (\ref{eq:con_correl_definition}). The net variance
is therefore given as
\begin{equation}
\sigma^{2}(\delta)\equiv\left[\left(N-\left[N\right]\right)^{2}\right]=\left[N\right]+\Delta_{{\rm scc}}(\delta),\label{eq:all_variance}\end{equation}
which is again an application of equation (7.66) in \citet{1993ppc..book.....P}
to the conditional cases we consider. 
This also suggests that the
true PDF deviates from the pure Poisson statistics, and the
super-Poissonian PDF
suggested by \citet{Saslaw1984}, given by
\begin{equation}
P_{{\rm
    cell}}(N)=\frac{\left[N\right]}{N!}e^{-\left[N\right](1-\beta)-N\beta}(1-\beta)\left([N](1-\beta)+N\beta\right)^{N-1},
\label{eq:pdf_superPoi}
\end{equation}
shows
excellent agreement with e.g. the distribution of N-body haloes of
$M=10^{10-11}\,M_{\odot}$ (\citealt{Neyrinck2014}). Here
$\beta\equiv 1-\sqrt{\sigma^{2}(\delta)/[N]}$ represents the degree of super-Poissonianity.

Sometimes, we may only be interested in those cells that contain at least one
halo. Quantifying this might be useful when haloes are rare, such that not all
the cells with given $\delta$ are occupied by these haloes. It is therefore
useful to have the conditional probability that there are $N$ haloes
in the cell (with $\delta$ and $V_{{\rm cell}}$) once a halo is
found in that cell (let us denote these cells by {}``active cells'').
This requires re-normalizing the PDF%
\begin{eqnarray}
P_{{\rm cell}}(N|N\ge1)&\equiv& P(N|\delta,\, V_{{\rm cell}};\,N\ge1) \nonumber \\
&=&\frac{P_{{\rm cell}}(N)}{P(N\ge1|\delta,\, V_{{\rm cell}})} \nonumber \\
&=&\frac{P_{{\rm cell}}(N)}{1-P(N=0|\delta,\, V_{{\rm cell}})} \nonumber \\
&=&\frac{P_{{\rm cell}}(N)}{1-e^{-\left[N\right](1-\beta)}},
\label{eq:concon_pdf_N}
\end{eqnarray}
where in the last equality we used equation (\ref{eq:pdf_superPoi}). 
The mean value of $N$ inside {}``active'' cells will then be given
by 
\begin{equation}
\left[N\right]_{a}\equiv\sum_{N=1}^{\infty}NP_{{\rm
    cell}}(N|N\ge1)=\frac{\left[N\right]}{1-e^{-\left[N\right](1-\beta)}},
\label{eq:mean_pdf}
\end{equation}
which should be used as the estimator of the mean value. Two 
limiting cases are noteworthy. First, when $\left[N\right]\ll1$,
$P_{{\rm cell}}(N|N\ge1)$ can be approximated as
\begin{equation}
P_{{\rm cell}}(N|N\ge1)\simeq
\frac{1}{N!}e^{-N\beta}(N\beta)^{N-1},
\label{eq:concon_pdf_smallN}
\end{equation}
which is no longer dependent on $[N]$.
In the other extreme, $[N]\gg1$, $P_{{\rm cell}}(N|N\ge1)=P_{{\rm
    cell}}(N)$.

Similarly, we use the same renormalization to determine the collapsed
fraction inside active cells:
\begin{equation}
\left[f_{c}(\delta)\right]_{a}=\frac{\left[f_{c}(\delta)\right]}{1-e^{-[N](1-\beta)}}.\label{eq:concon_fcoll}\end{equation}
When $\left[N\right]\ll1$, as the mass function is biased toward
the least massive haloes, $\left[f_{c}\right]_{a}\simeq M_{{\rm min}}/M_{{\rm cell}}=M_{{\rm min}}\rho_{0}^{-1}V_{{\rm cell}}^{-1}(1+\delta)^{-1}$.
When $\left[N\right]\gg1$, $\left[f_{c}\right]_{a}$ trivially converges
to $\left[f_{c}\right]$.

Note that $\left[N\right]$ can be smaller than 1. This does not mean
that we will find a fractional, less-than-unity number of haloes on
average, which is simply unphysical. This means instead, assuming
ergodicity, that 
\begin{eqnarray}
\left[N\right]&=&\frac{{\rm total\, number\, of\, haloes\, found\, in\, all\,
    cells\, with\, \delta}}{{\rm total\, number\, of\, cells\, with\,
    \delta}} \nonumber \\
&\approx&\frac{{\rm number\, of\, active\, cells\, with\,
    \delta}}{{\rm total\, number\, of\, cells\, with\, \delta}},
\label{eq:meaning_smallN}
\end{eqnarray}
where the approximation is made possible due to the fact that when
$\left[N\right]\ll1$, the PDF $P_{{\rm cell}}(N|N\ge1)$ is peaked
at $N=1$.

\section{Results}

\label{sec:Result}


\subsection{Mean Unconditional Halo Mass Function}

\label{sub:Result:Halo-Mass-Function}

The mean, unconditional halo mass functions at both high and low redshifts
were recently discussed in detail in \citet{Watson2014}, much of
it based on the same simulations as the current work. Therefore, we
will only summarize a selection of the mean mass function properties
that are most relevant here.

In Fig.~\ref{fig:meanMF} we show the mass functions in the mass
range $M\ge10^{5}\, M_{\odot}$ at selected redshifts based on the
$L_{{\rm box}}=20/h\,{\rm Mpc}$ and $6.3/h\,{\rm Mpc}$ simulations,
together with PS and ST analytical mass functions. The actual quantities
plotted are halo number densities $\Delta n\equiv\int_{M_{1}}^{M_{2}}(dn/dM)dM$,
integrated over equal-size logarithmic mass bins. The last mass bin
includes all haloes with mass $M\ge10^{9}\, M_{\odot}$. The two simulated
mass functions show excellent agreement with each other, except for
the high mass end, where the mass function is truncated due to finite
volume. This agreement indicates the consistency of the N-body
simulation over varying box size.

Compared to the analytical expressions, our N-body mass functions
are in better agreement with ST than PS mass functions. The agreement
with ST at all redshifts is within $\sim25\%$ for $M=[10^{5}-10^{6}]\, M_{\odot}$,
the haloes in which range numerically dominate the minihalo population.
At very high redshifts, $z\gtrsim20$, ST mass function slightly over-predicts
halo population at $M=[10^{5}-10^{5.5}]\, M_{\odot}$ and under-predicts
halo abundance at $M\ge10^{6}\, M_{\odot}$, with tendency to deviate
increasingly as $M$ increases, while at relatively low redshifts,
over-prediction occurs at $M=[10^{5.5}-10^{6}]\, M_{\odot}$. As discussed
in \citet{Watson2014} these differences are partly due to our usage
of a halo finder based on spherical overdensity instead of the friends-of-friends
one used by ST, and also to the limitations of the ST fit which was
based on low-redshift data and relatively small simulations. In contrast,
the classical PS mass function gives a poor fit to N-body minihalo
data at all redshifts, severely under-predicting the abundance of
rare ($\nu=\delta_{c}/\sigma_{M}\gg1$) haloes and over-predicting
the abundance of $\nu\ll1$ haloes. Only for the most common ($\nu\approx1$)
haloes PS is a more reasonable approximation (and also agrees with
ST).

Assuming that the prescription for the conditional mass function (linking
eq.~\ref{eq:boost} with unconditional mass function) provides a
correct theoretical framework, one may expect that a good fit to unconditional
mass function will also provide a good fit to conditional mass function
when combined with eq. (\ref{eq:boost}). Therefore, we can expect
that $\left(dn/dM\right)_{{\rm N-body},\,{\rm b}}$ will be
the best fit to the mean conditional mass function from the simulations,
and $\left(dn/dM\right)_{{\rm ST},\,{\rm b}}$ will also be
a good fit, while $\left(dn/dM\right)_{{\rm PS},\, b}$ will
be a poor fit. We will test this expectation in Section \ref{sub:Result:mean}.

\subsection{Mean biased halo mass function}

\label{sub:Result:mean}

We now show how the mean, conditional mass functions of N-body haloes
behave in terms of $\delta$, and compare this to the modelling predictions
based on the different mass functions, $\left(dn/dM\right)_{{\rm PS},\, b}$,
$\left(dn/dM\right)_{{\rm ST},\,{\rm b}}$ and $\left(dn/dM\right)_{{\rm N-body},\,{\rm b}}$.
We also compare these to the model based on the linear bias. The stochasticity
in this relation will be treated in \S~\ref{sub:Result:scatter}.

\subsubsection{Minihaloes}

\label{sub:Minihaloes}

Minihaloes are usually defined by their hydrodynamical properties.
Their minimum mass is the cosmic Jeans mass determined by the mean
IGM temperature, and their maximum mass is the mass of haloes whose
virial temperature is about $10^{4}\,{\rm K}$. While this is the
general definition, the uncertainty of the mean IGM temperature at
high redshift makes the definition of the minimum mass somewhat uncertain.
In this work we instead take their mass to be in a fixed range $M=[10^{5}-10^{8}]\, M_{\odot}$,
which is of more direct use to N-body data at fixed mass resolution.
Both $6.3/h\,{\rm Mpc}$ and $20/h\,{\rm Mpc}$ boxes resolve haloes
down to $M=10^{5}\, M_{\odot}$. The latter simulation thus provides
a better statistics by encompassing a volume 32 times as large as
that of the former one.

\begin{figure*}
\includegraphics[width=80mm]{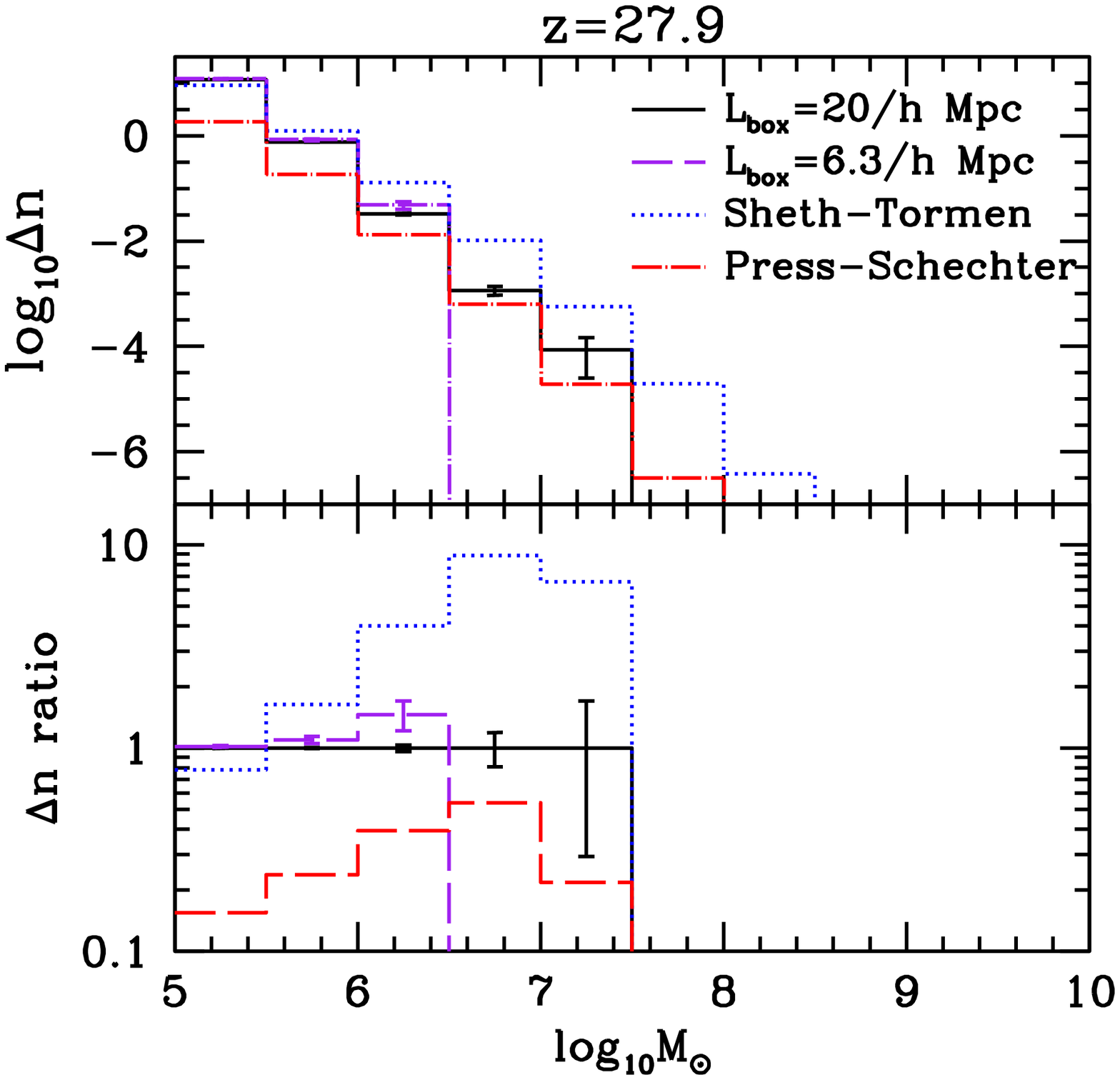}\includegraphics[width=80mm]{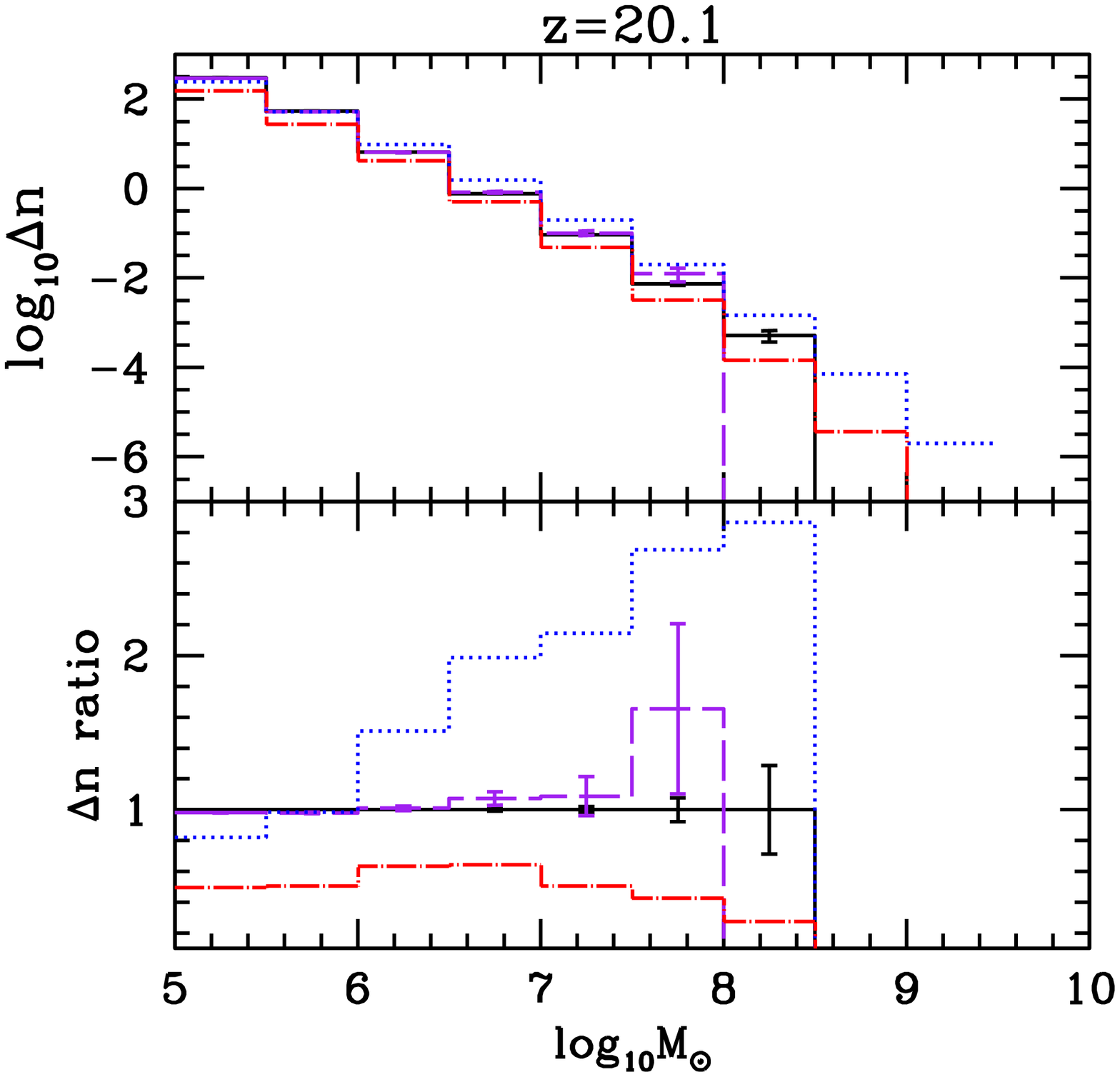}

\includegraphics[width=80mm]{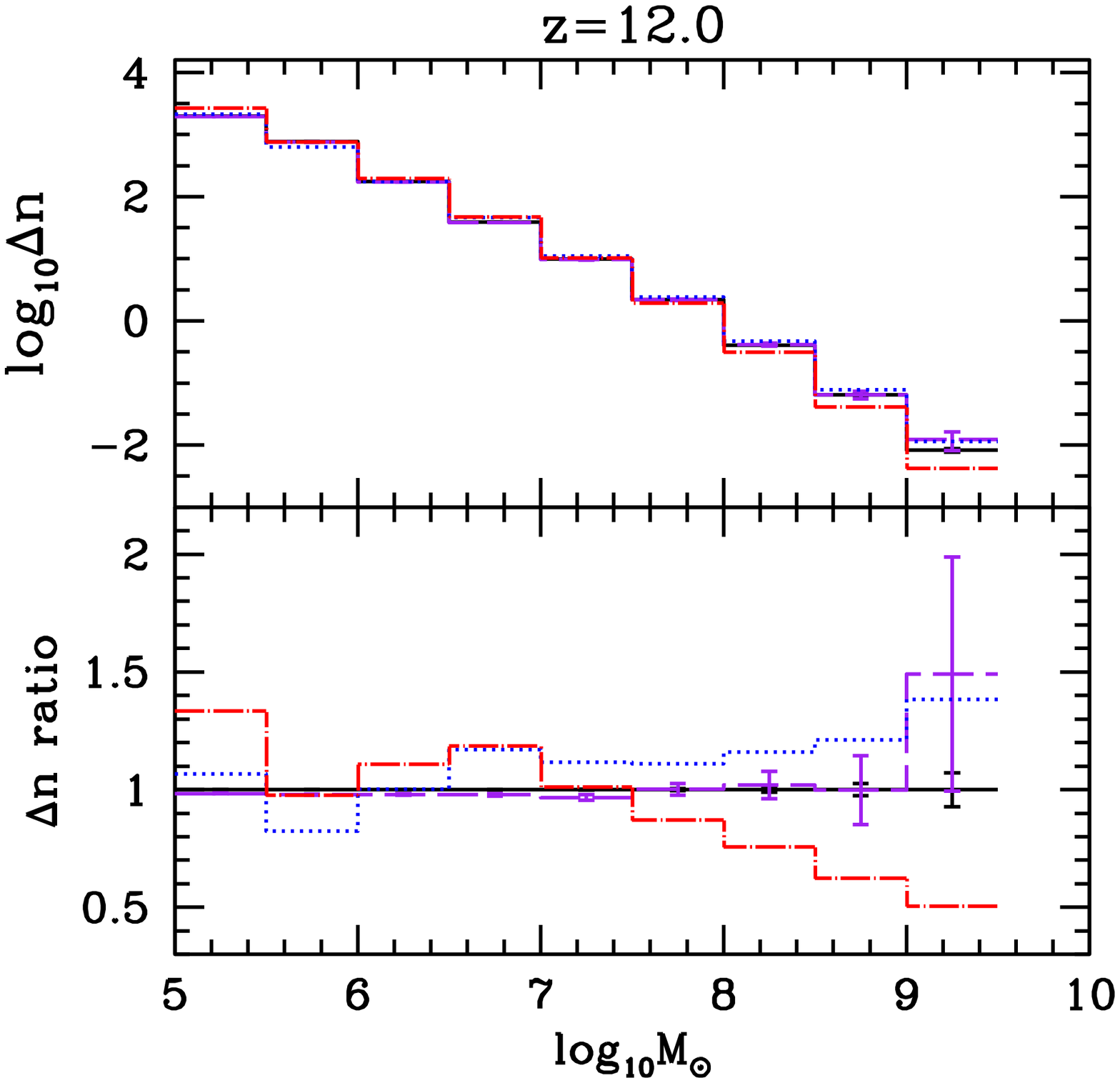}\includegraphics[width=80mm]{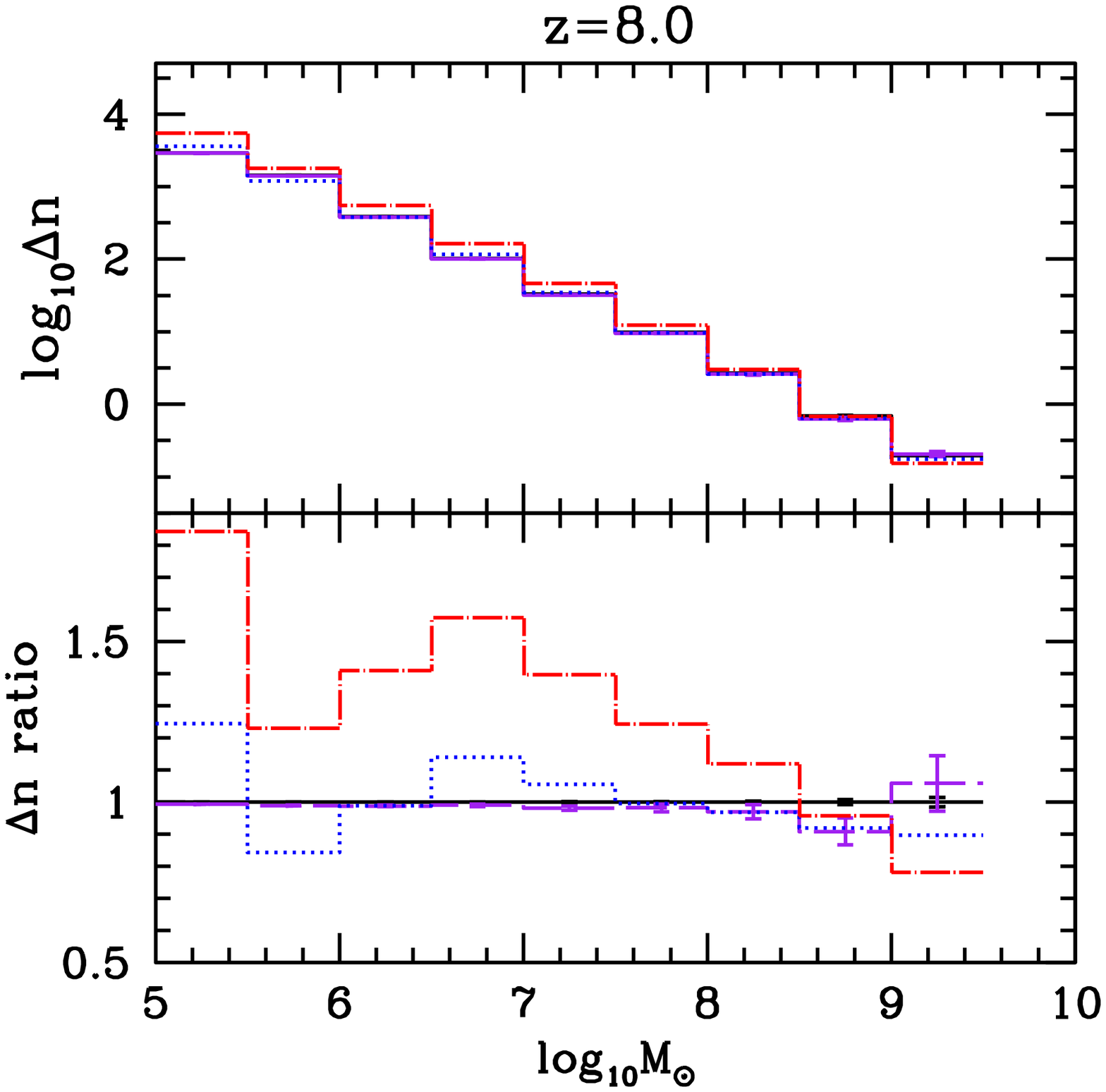}

\caption{Unconditional, mean mass functions of MHs found in small-box N-body
simulations. Plotted are the mass functions (each top panel) inside $20/h$~Mpc (black,
solid) and $6/h$~Mpc (purple, dashed) boxes, along with
the Sheth-Tormen (blue, dotted) and Press-Schechter (red, dot-dashed) mass
functions, all integrated 
over equal-size logarithmic mass bins, and the ratios of these mass
functions (each bottom
panel with the same line types) to the mass
function inside the $20/h$ Mpc box. The error bars represent
$1\sigma$ standard deviation in each mass bin.}
\label{fig:meanMF}
\end{figure*}

\begin{figure*}
\includegraphics[width=80mm]{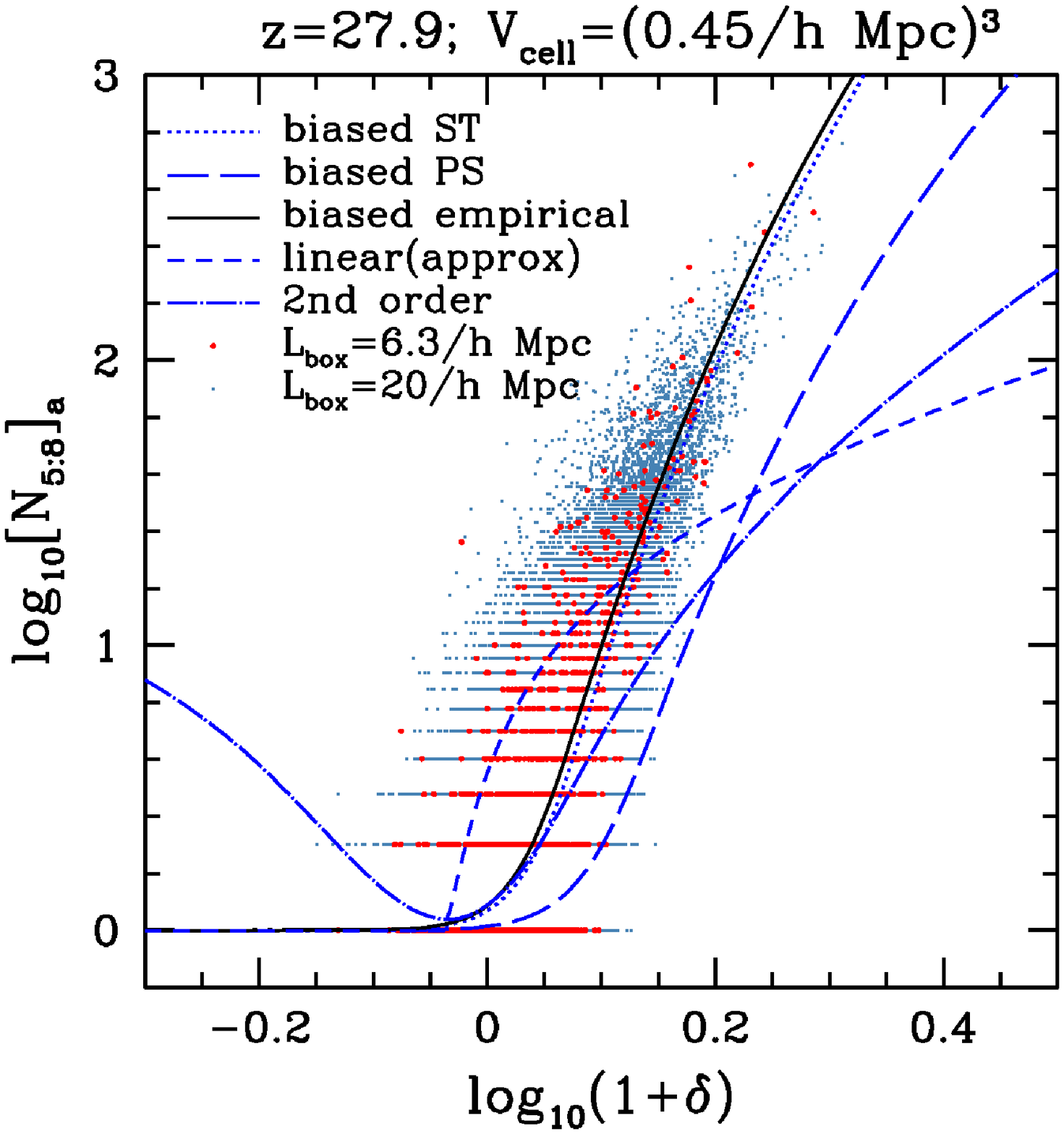}\includegraphics[width=80mm]{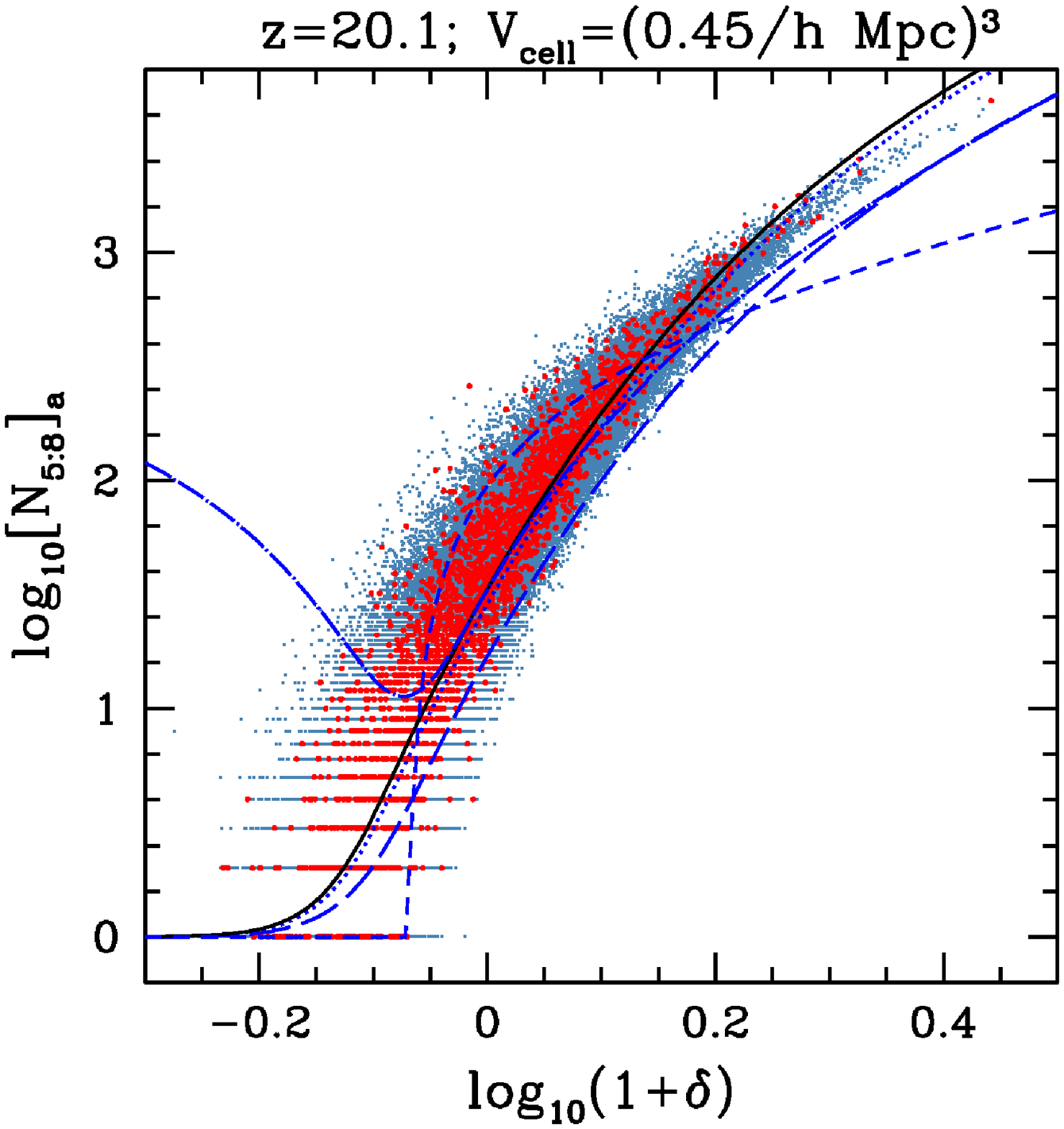}

\includegraphics[width=80mm]{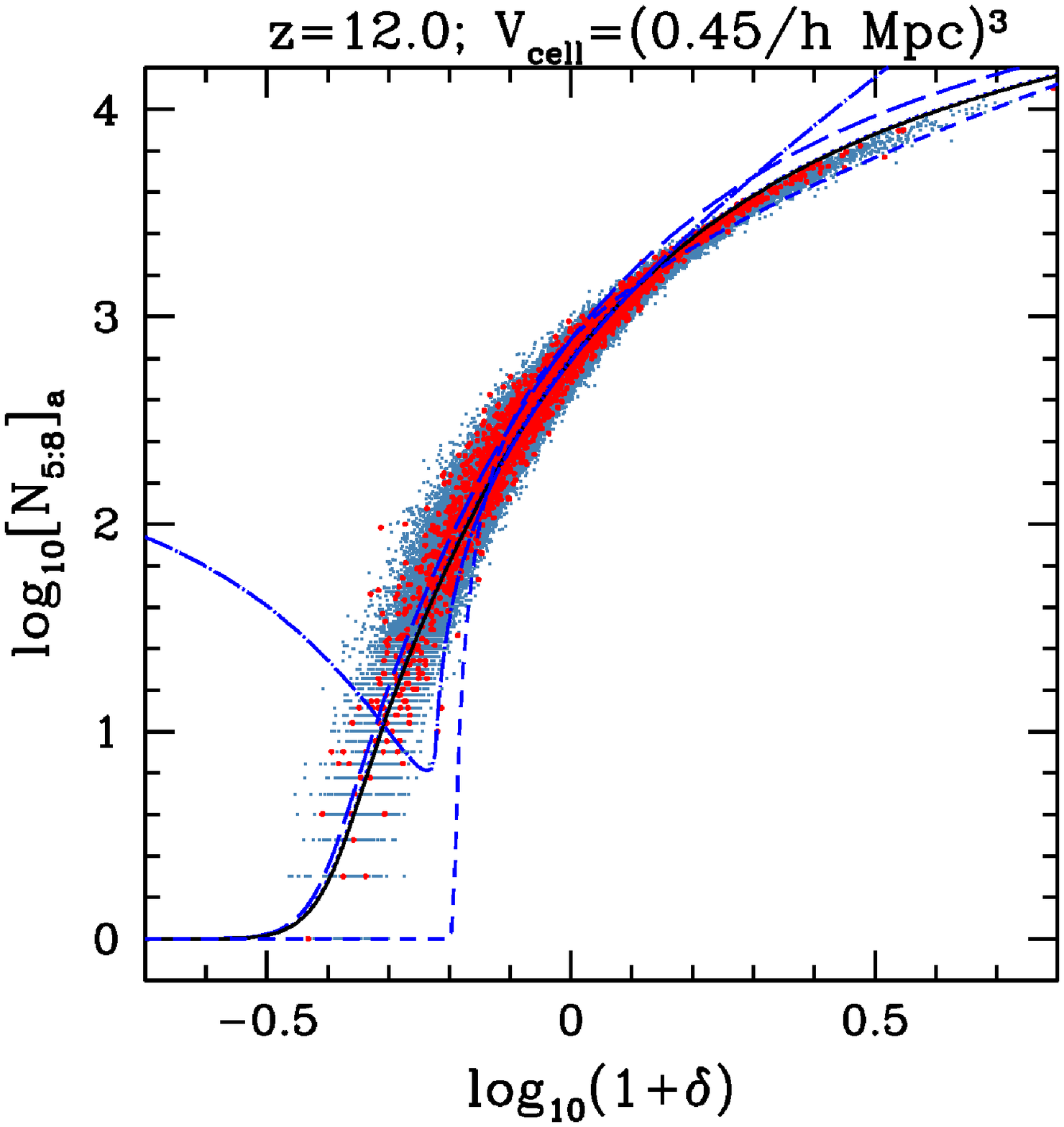}\includegraphics[width=80mm]{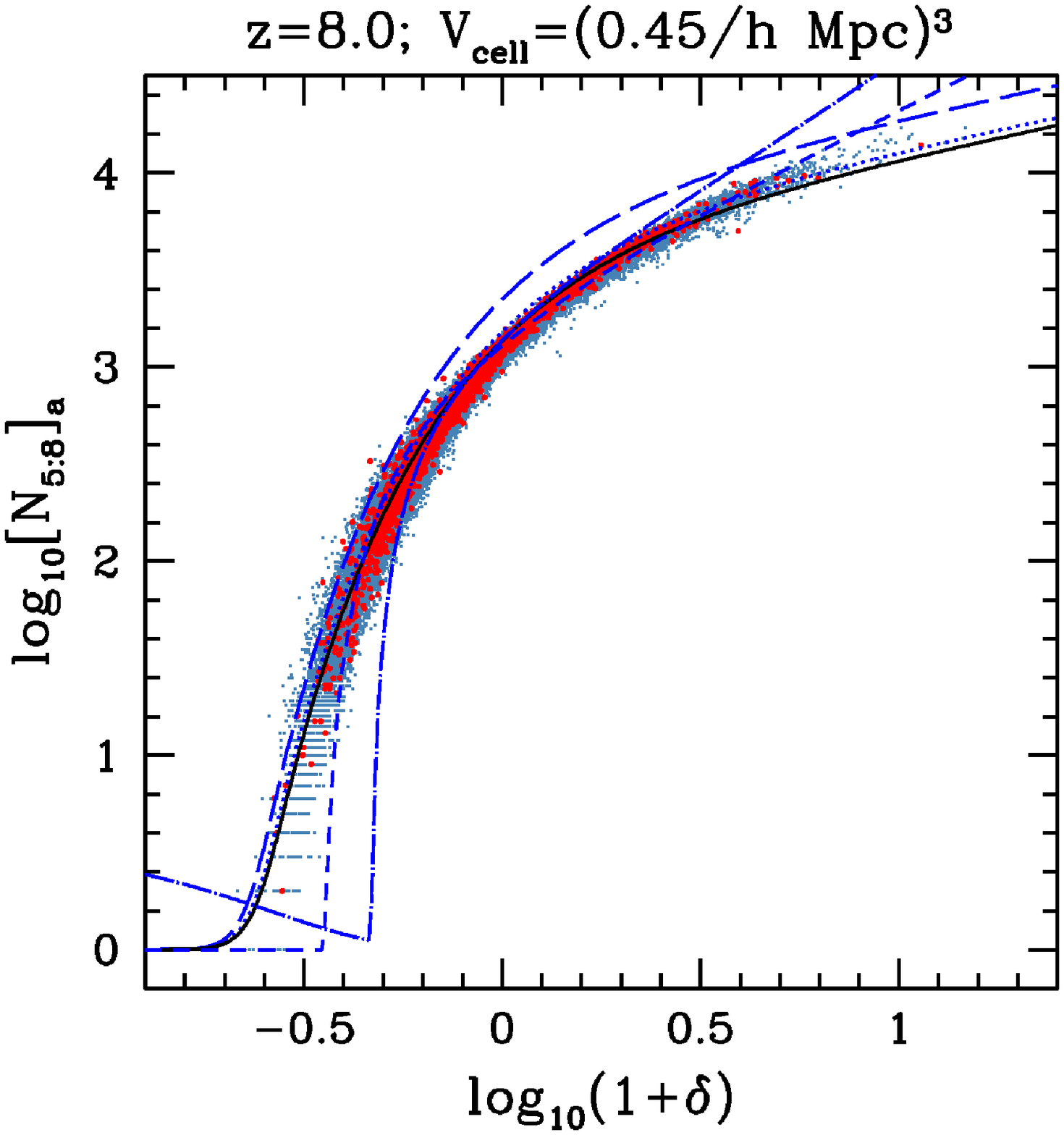}

\caption{Correlation between the number of minihaloes $N_{5:8}$ and the
cell overdensity $\delta$, when the Eulerian volume of the cell is
chosen to be $(0.45/h\,{\rm Mpc})^{3}$. Data points are from N-body
simulations in the 6.3/h Mpc box (red, larger dot) and 20/h Mpc box
(slate blue, smaller
dot), which are sampled by $14^{3}$ and $45^{3}$ cells, respectively.
Theoretical predictions for active cells (eq. \ref{eq:mean_pdf}) based
on $\left(dn/dM\right)_{{\rm N-body},\, b}$
(solid, black; eq.~\ref{eq:Nbody-bias}),
$\left(dn/dM\right)_{{\rm ST},\, b}$ 
(dotted, blue; eq.~\ref{eq:ST-bias}), 
$\left(dn/dM\right)_{{\rm PS},\, b}$ 
(long-dashed, blue; eq.~\ref{eq:NL_lin_dndM}),
the one by the linear
bias approximation 
without the 0-point offset $B_{0}$ defined in 
Section~\ref{sub:Validity-of-Linear}
(short-dashed, blue; eq.~\ref{eq:bias_linear})
combined with $\left(dn/dM\right)_{{\rm N-body},\, b}$  
and
the one by the 2nd order approximation
with $B_0$
(dot-dashed, blue; equations~\ref{eq:2nd-expand}-\ref{eq:q_define})
also combined with $\left(dn/dM\right)_{{\rm N-body},\, b}$
are plotted for comparison. }
\label{fig:meannumber-mini-6.3-14}
\end{figure*}

\begin{figure*}
\includegraphics[width=80mm]{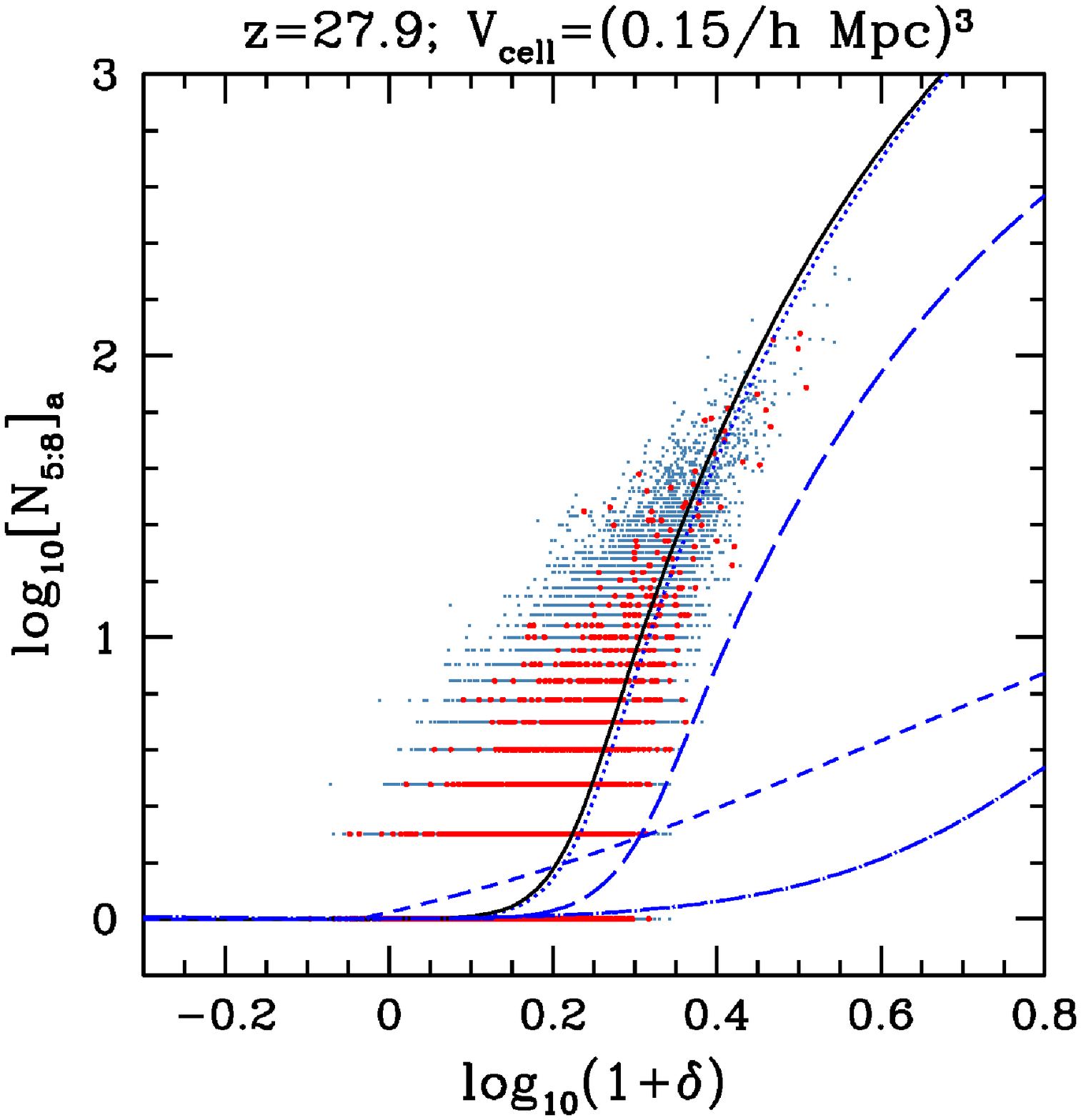}\includegraphics[width=80mm]{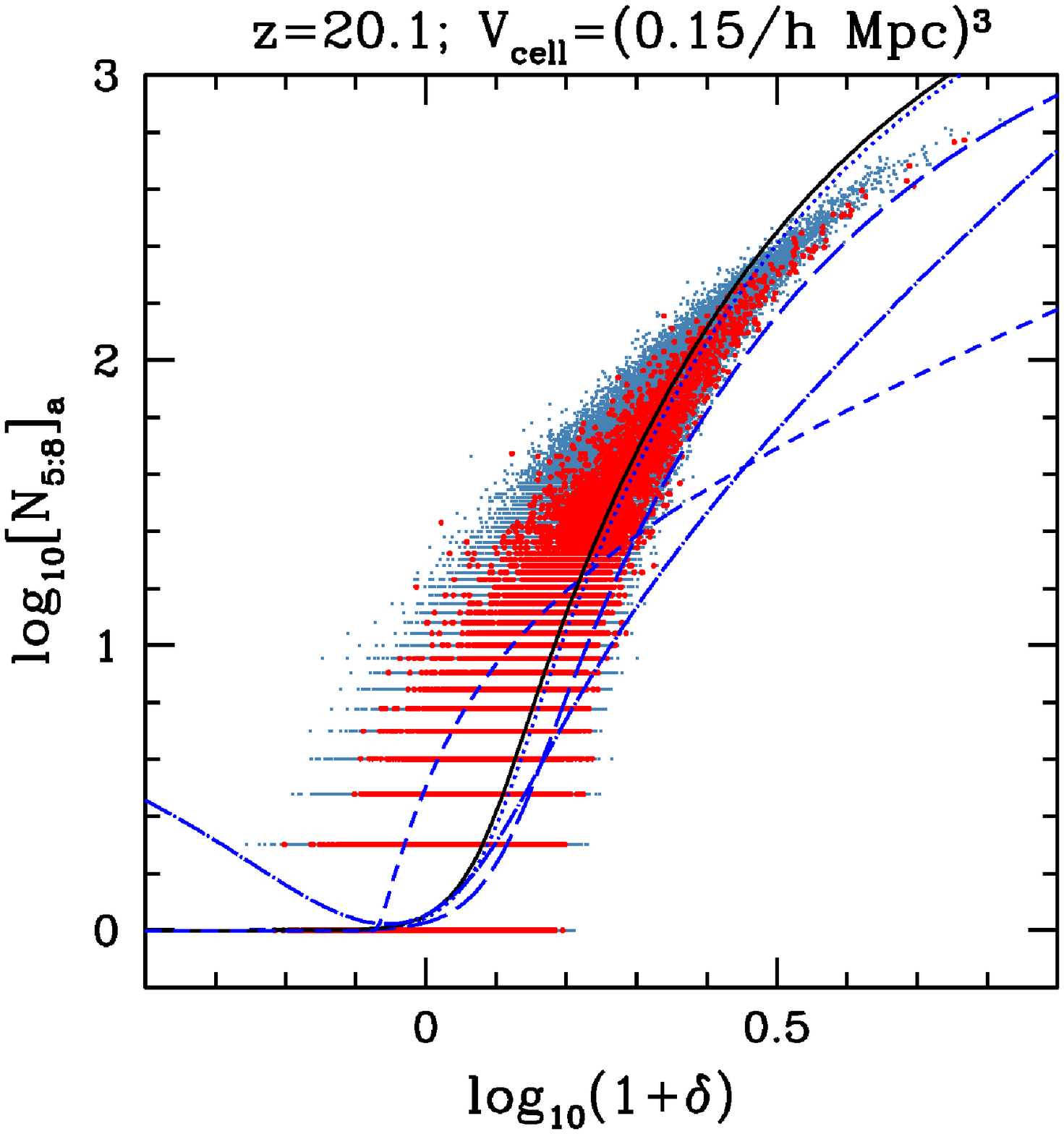}
\includegraphics[width=80mm]{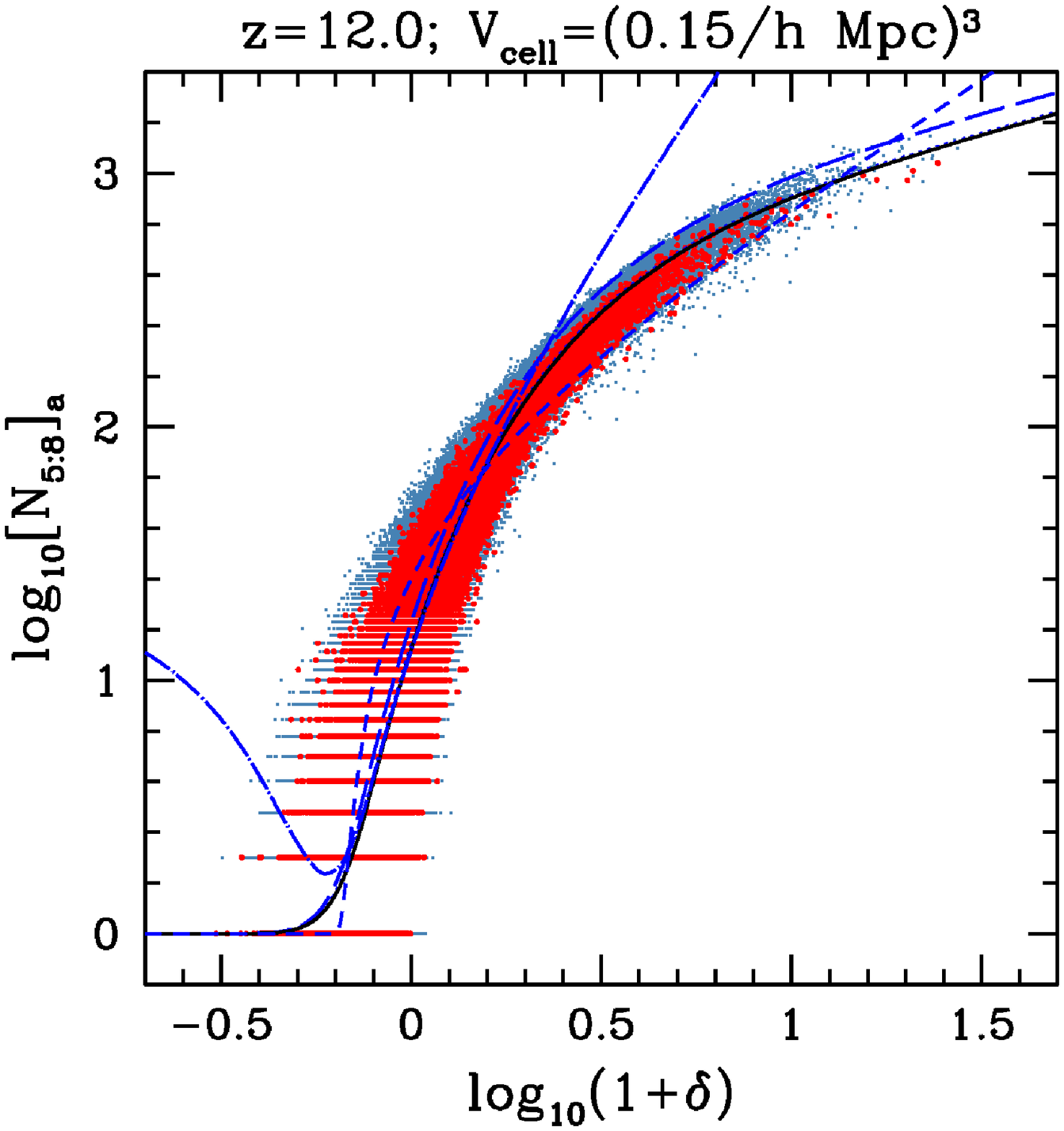}\includegraphics[width=80mm]{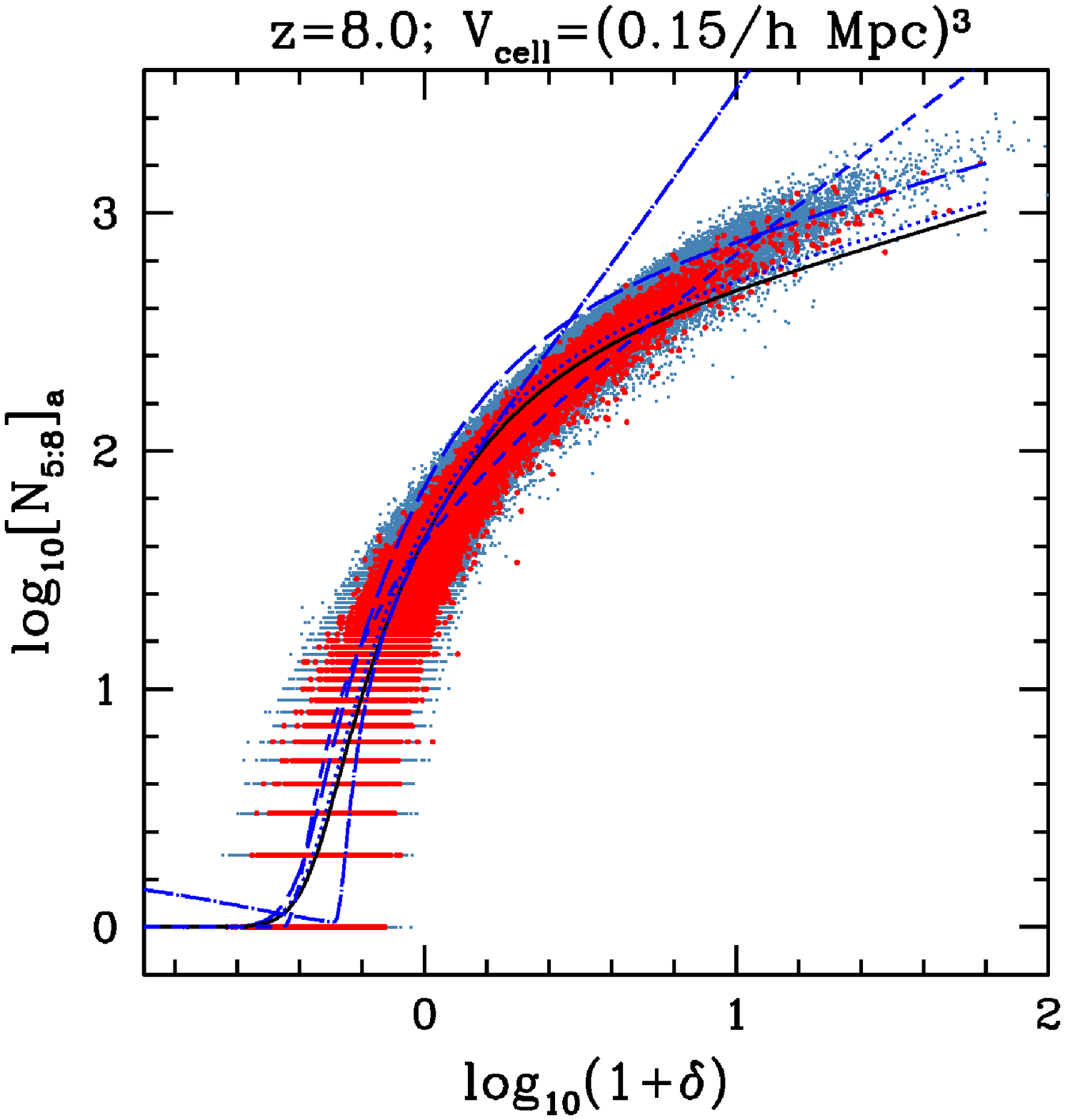}

\caption{Same as Fig. \ref{fig:meannumber-mini-6.3-14}, except that the volume
of the cell is now $(0.15/h\,{\rm Mpc})^{3}$. The 6.3/h Mpc box (red
circle) and 20/h Mpc box (green dot) are sampled by $44^{3}$ and
$135^{3}$ cells, respectively.}
\label{fig:meannumber-mini-6.3-44}
\end{figure*}

We first examine how well the models based on the analytical mass
function fits match the N-body
data. 
Figs~\ref{fig:meannumber-mini-6.3-14} and \ref{fig:meannumber-mini-6.3-44} 
show the analytical estimates and N-body data on the total number
of minihaloes $\left[N_{5:8}\right]_{a}$ 
under different Eulerian cell sizes. We find that the
numerical data from the two simulation volumes is in excellent agreement
and that$\left(dn/dM\right)_{{\rm ST},\,{\rm b}}$ and $\left(dn/dM\right)_{{\rm N-body},\,{\rm b}}$
fit the N-body data well over almost the entire range of $\delta$
and $z$, while $\left(dn/dM\right)_{{\rm PS},\,{\rm b}}$
and the linear relation $\delta_{h}=b_{\rm lin}\delta$ both
provide poor fits to the data in 
general. Even though $\left[N_{5:8}\right]_{a}$ and $\left[f_{c,\,5:8}\right]_{a}$
are integral quantities, given that smallest-mass haloes numerically
dominate the halo population, both the data and semi-analytical estimates
reflect predominantly the low-mass end. 
Note that as seen in Fig.~\ref{fig:meanMF}, $\left(dn/dM\right)_{\rm
  ST}$ agrees well with $\left(dn/dM\right)_{\rm N-body}$ in the low
mass-end, and this is the reason why 
$\left(dn/dM\right)_{{\rm ST},\,{\rm b}}$ provides a good fit. If we
focused on the 
high-mass end only, $\left(dn/dM\right)_{{\rm ST},\,{\rm b}}$ would be
a very poor fit to the observed bias, because the average ST mass function
$\left(dn/dM\right)_{\rm ST}$ has large discrepancy from the actual
N-body data for e.g. $M\ge 10^{7} M_{\odot}$.
In contrast, the collapsed
fraction in haloes $\left[f_{c,\,5:8}\right]_{a}$ (Figs~1 and 2 of the
Supplementary Material) is a mass-weighted quantity
and thus reflects the high mass end better than does $\left[N_{5:8}\right]_{a}$,
but the rapid exponential cutoff of the mean halo mass function $dn/dM$
at increasing $M$ still moderates the contribution from the high-mass
haloes. The similarity between $\left(dn/dM\right)_{{\rm ST},\,{\rm b}}$
and $\left(dn/dM\right)_{{\rm N-body},\,{\rm b}}$ reflect
the simple fact that the unconditional mass functions, 
$\left(dn/dM\right)_{{\rm ST}}$
and $\left(dn/dM\right)_{{\rm N-body}}$, are similar around
the low-mass end.

As the cell size shrinks, however, some discrepancy appears at
high $\delta$ regime. Both $\left(dn/dM\right)_{{\rm ST},\,{\rm b}}$
and $\left(dn/dM\right)_{{\rm N-body},\,{\rm b}}$ predictions
overestimate the N-body data substantially at $\delta\gtrsim1.5$,
when the volume of the cell has shrunken from $(0.45/h\,{\rm Mpc})^{3}$
to $(0.15/h\,{\rm Mpc})^{3}$: see Fig.~\ref{fig:meannumber-mini-6.3-44}
At this point, where $\delta$
approaches the overdensity criterion for halo identification, we suspect
that this could be a symptom of extreme nonlinearity: the mean mass
of the cell, $M_{{\rm cell}}=3.8\times10^{8}\, M_{\odot}$, is small
enough to be comparable to the high-mass end of minihaloes, or
$10^{8}M_{\odot}$. 

In summary, unless the cell is too small, and thus potentially quite
nonlinear, the mean nonlinear bias of N-body minihaloes at high redshifts
can be explained well by the simple hybrid prescriptions 
$\left(dn/dM\right)_{{\rm N-body},\,{\rm b}}$
and $\left(dn/dM\right)_{{\rm ST},\,{\rm b}}$. In contrast,
at high redshifts, 
the linear relation $\delta_{h}\propto \delta$ deviates too much
from the N-body 
minihalo data to be of much practical use at least under the filtering
scales of $\lesssim \rm Mpc$. The disagreement of
$\left(dn/dM\right)_{{\rm PS},\,{\rm b}}$
with the N-body data is just as severe, and 
we expect that $\left(dn/dM\right)_{{\rm PS},\,{\rm b}}$ will
be useless regardless of the filtering scale, because the disagreement
is caused by the poorness of the mean PS mass function.
It is notable that the rarity 
of haloes at high redshifts make the linear relation
fail even when $\left|\delta\right|\ll1$, which will be discussed
in much detail in Section \ref{sub:Validity-of-Linear}.

\subsubsection{Atomically-cooling haloes}
\label{sub:Atomhalo}

Atomically-cooling haloes (ACH hereafter) are named after the dominant
cooling mechanism of baryonic gas inside. Atomic line radiation can
cool primordial-composition gas to $T\simeq10^{4}\,{\rm K}$ from
its initially higher virial temperature. Star formation are believed
to occur inside these haloes as pre-existing metals or newly-formed
${\rm H}_{2}$ can further cool the gas down to $T\sim100\,{\rm K}$.
Therefore, ACHs are usually defined by their virial temperature: haloes
with $T\gtrsim10^{4}\,{\rm K}$. As this threshold virial temperature
roughly coincides with $M\simeq10^{8}\, M_{\odot}$, here we define
ACHs as those haloes with $M\ge10^{8}\, M_{\odot}$. The ACHs can be
grouped further into low-mass ACHs (LMACH), for which the gas pressure
of the photoheated IGM in an ionized patch prevented the halo from
capturing the gas it needs to form stars, and high-mass ACHs (HMACH),
for which gravity was strong enough to overcome this {}``Jeans-mass
filter'' and form stars even in the ionized patches. The dividing
line between LMACHs and HMACHs occurred roughly at $\sim10^{9}\, M_{\odot}$
(although the precise boundary value is still uncertain).

As our $114/h\,{\rm Mpc}$ box simulation resolves haloes of $M\ge10^{8}\, M_{\odot}$,
ACHs defined as above are fully identified. Even though the inner
structure of low-mass end haloes is not resolved near the resolution
limit (see Section \ref{sec:nbody}), for our considerations only
the number count of haloes matters, both for the mean halo bias and
stochasticity%
\footnote{As to be seen in \S~\ref{sub:Statistics} and \S~\ref{sub:Result:scatter},
the conditional halo correlation function determines the stochasticity.
The halo correlation function is composed of the 1-halo term and the
2-halo term, and the dominant contribution to stochasticity comes
from the 2-halo term. Therefore, it is not required
to fully resolve the halo structure in estimating the stochasticity.%
} and therefore our results are not affected by this.

\begin{figure*}
\includegraphics[width=80mm]{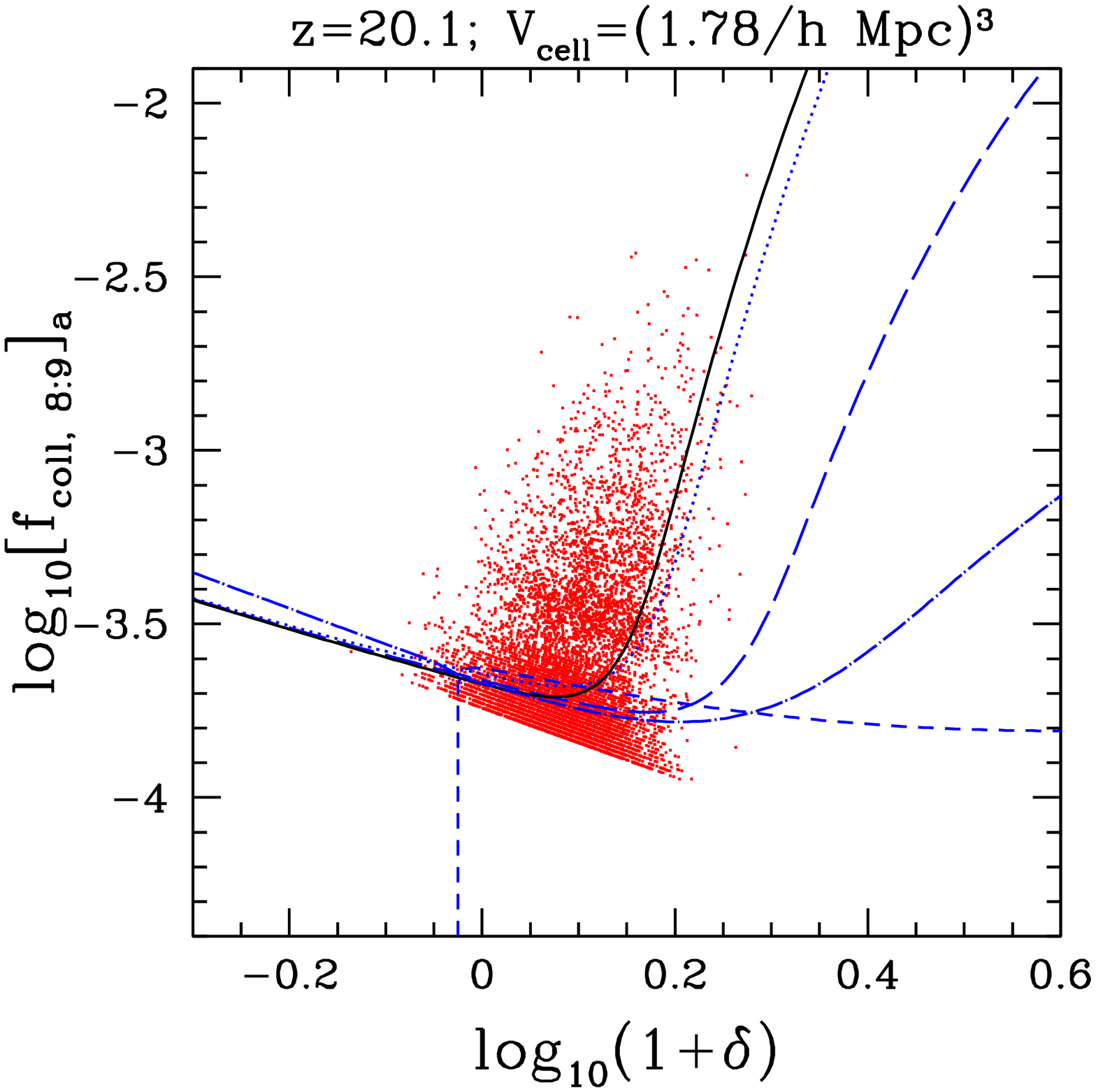}\includegraphics[width=80mm]{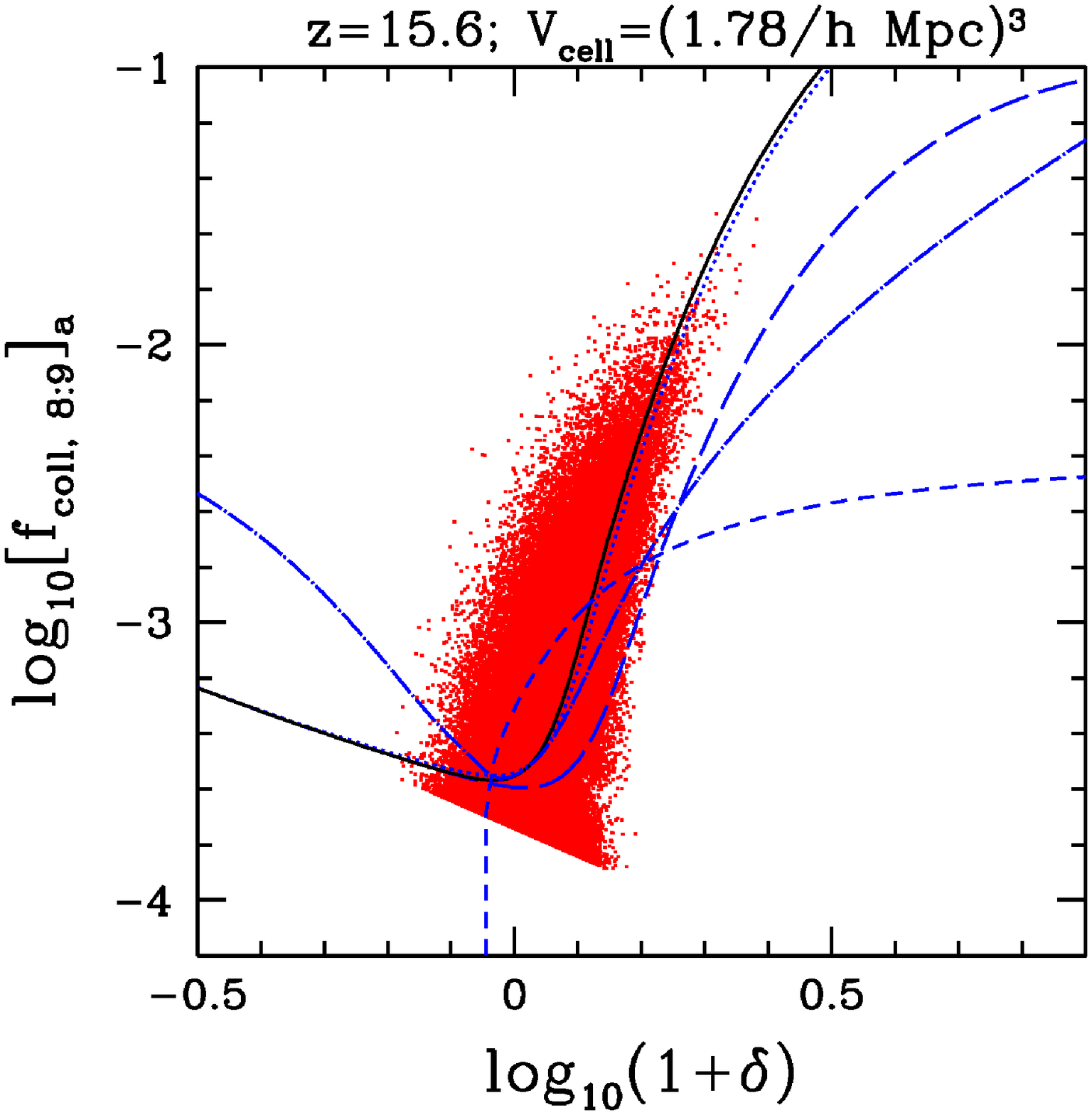}

\includegraphics[width=80mm]{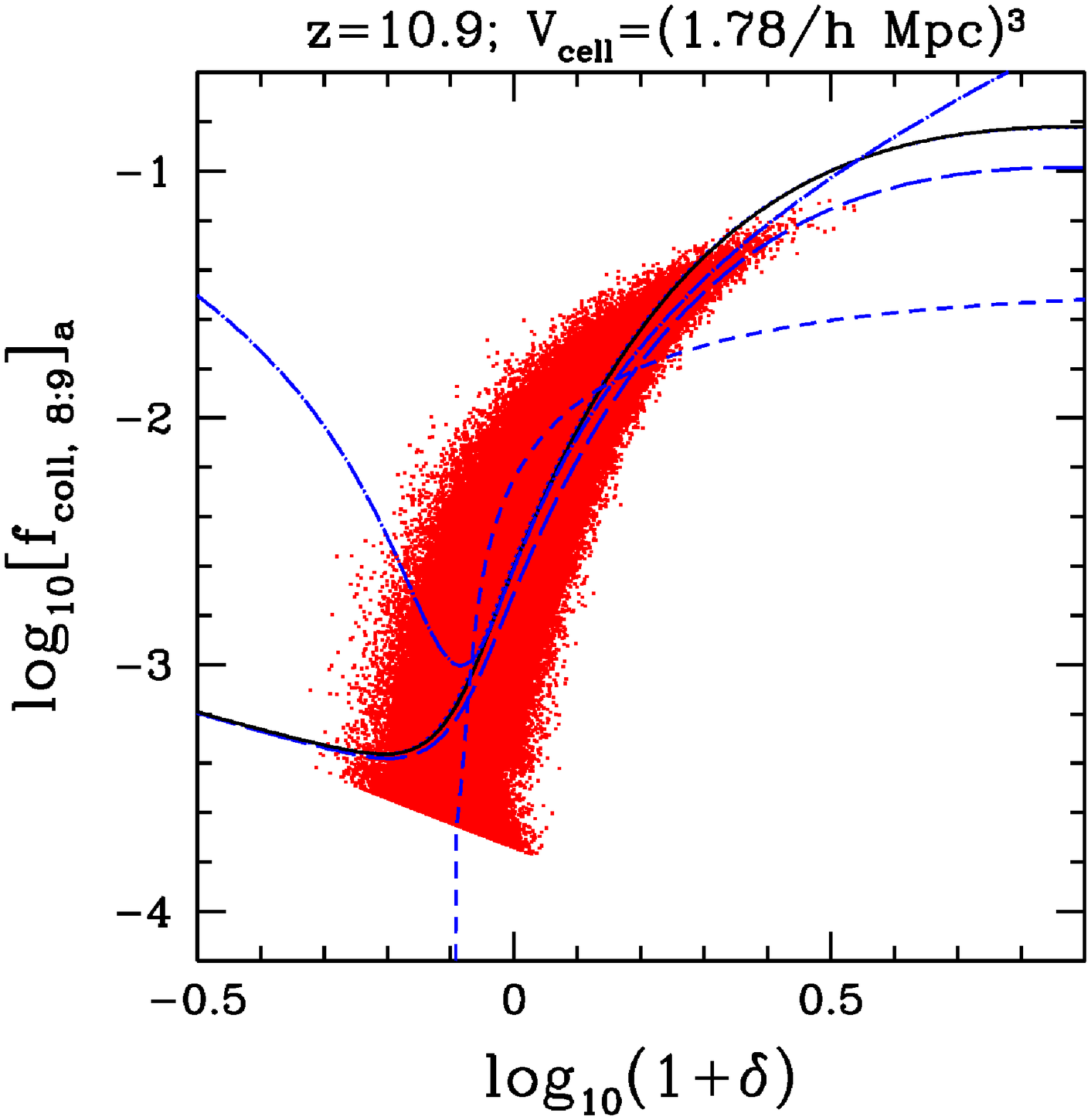}\includegraphics[width=80mm]{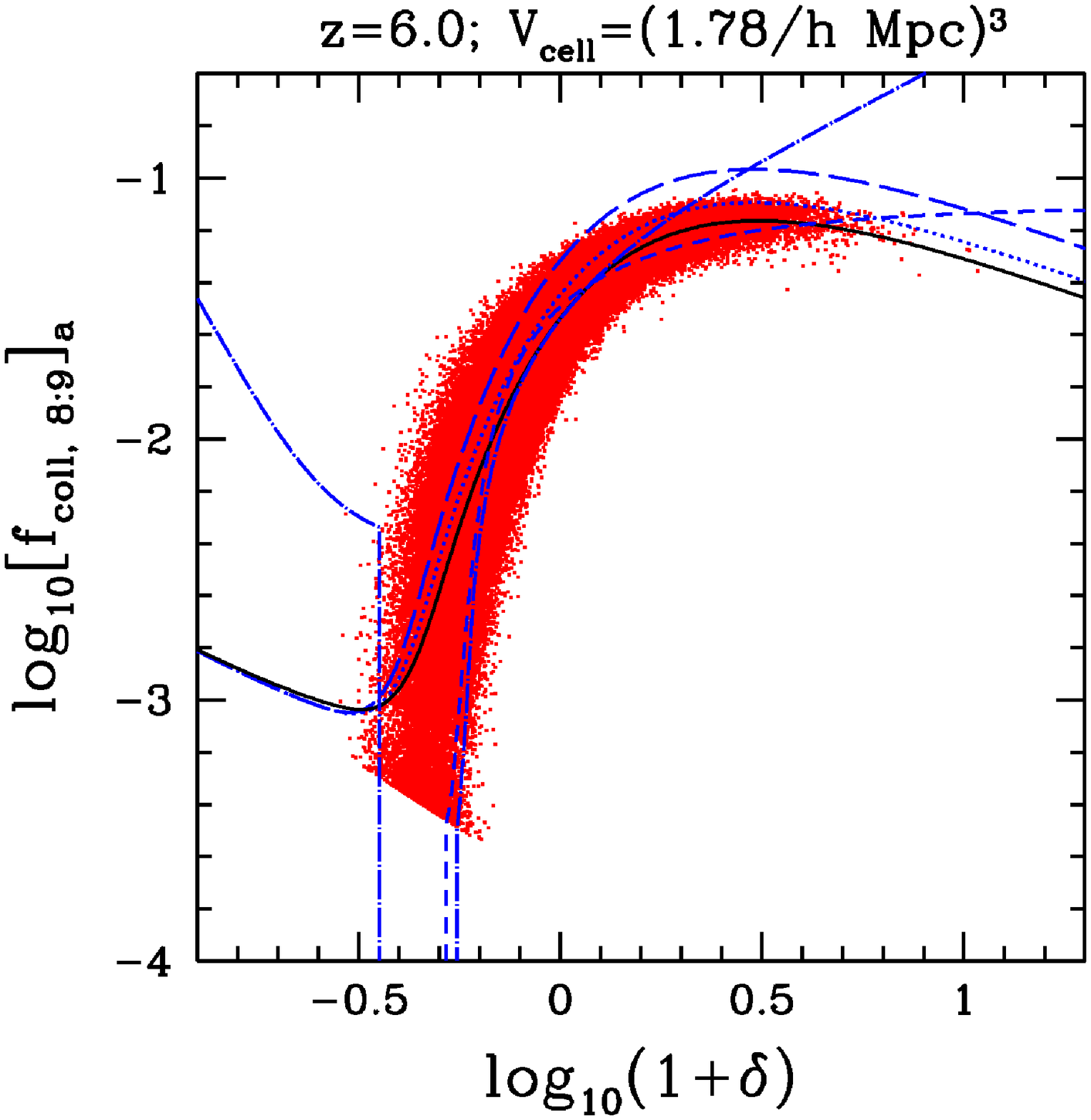}

\caption{Correlation between the fraction of mass collapsed ($f_{{\rm coll,\,8:9}}$)
into LMACHs ($M=10^{8}-10^{9}\, M_{\odot}$) and the cell overdensity
$\delta$ in the 114/h Mpc box, where the box is sampled by $64{}^{3}$
grid-cells. Conventions for plotting follow those of Fig. \ref{fig:meannumber-mini-6.3-14}, except for the data points
(red point).}
\label{fig:meanfcoll-LMACH-114-64}
\end{figure*}

\begin{figure*}
\includegraphics[width=80mm]{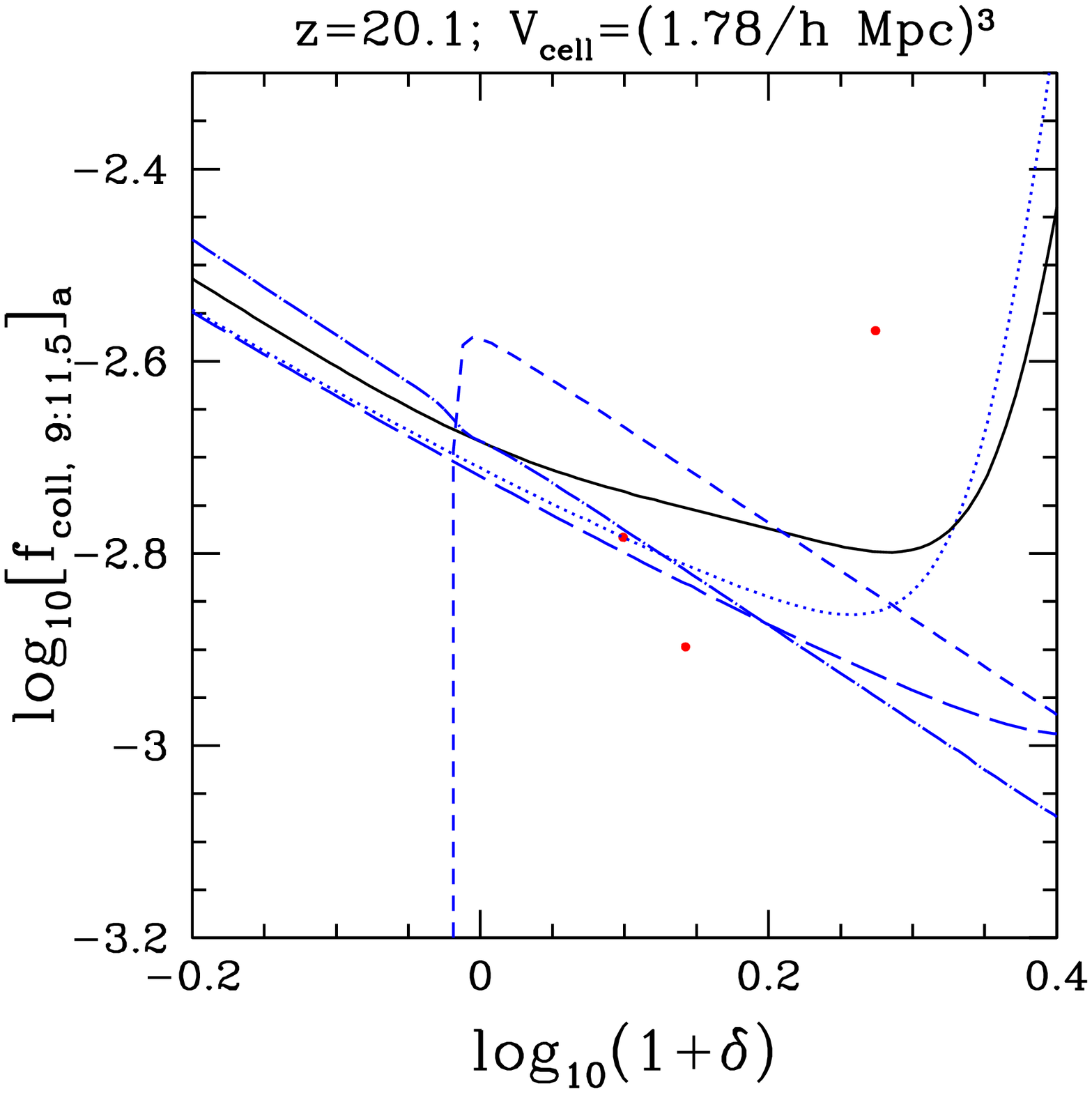}\includegraphics[width=80mm]{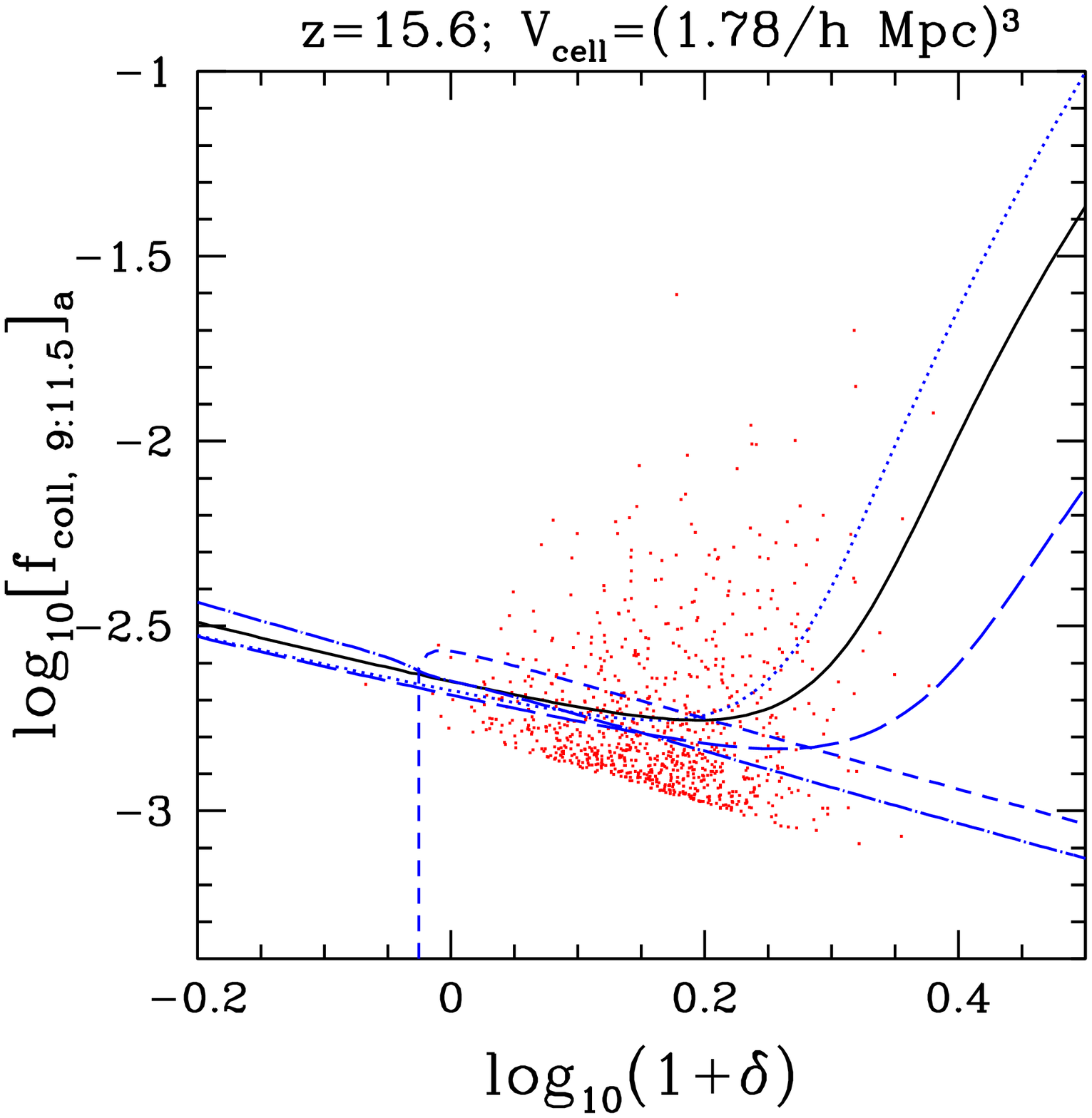}

\includegraphics[width=80mm]{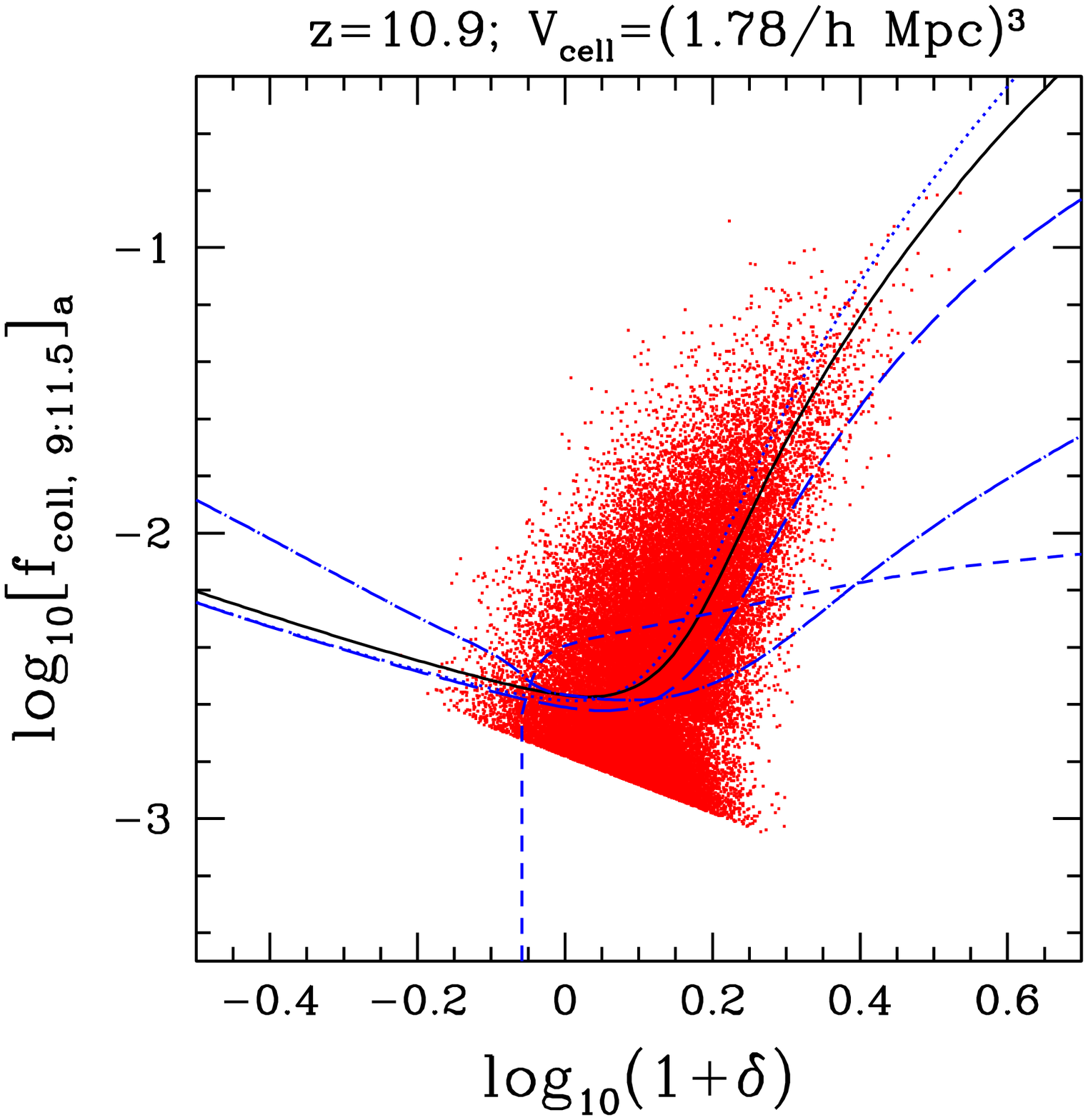}\includegraphics[width=80mm]{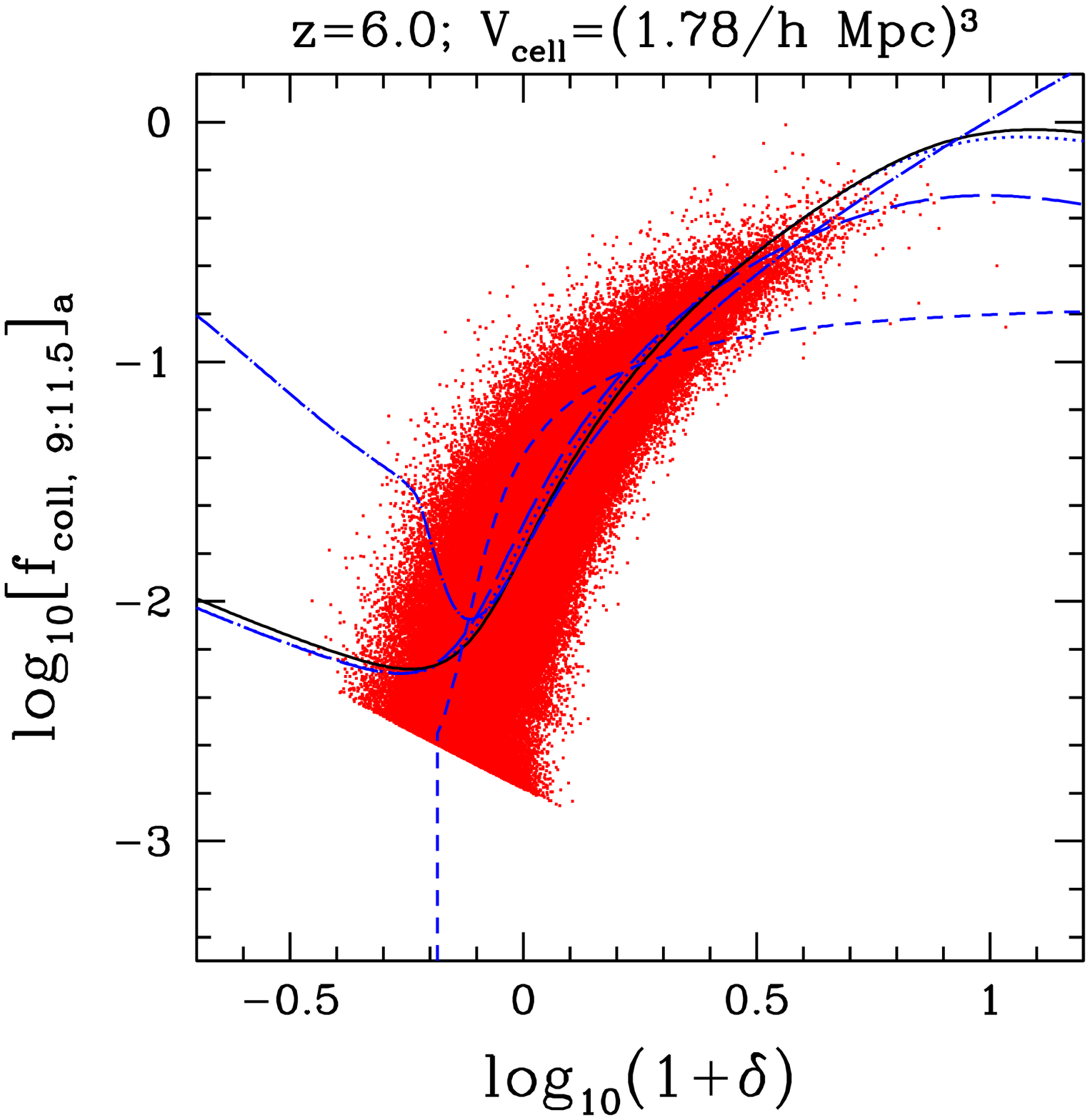}

\caption{Correlation between the fraction of mass collapsed 
($f_{{\rm coll,\,9:11.5}}$) 
into HMACHs ($M=10^{9}-10^{11.5}\, M_{\odot}$), where the maximum
mass is roughly the mass of a cell) and the cell overdensity $\delta$
in the 114/h Mpc box, where the box is sampled by $64{}^{3}$ grid-cells.
Conventions for plotting follow those of Fig. \ref{fig:meannumber-mini-6.3-14}, except for the data points
(red point).}
\label{fig:meanfcoll-HMACH-114-64}
\end{figure*}

We choose two filtering scales, $114/h/64=1.78/h\,{\rm Mpc}$ and
$114/h/32=3.56/h\,{\rm Mpc}$. While these 
choices are somewhat arbitrary, we increased the filtering scales for
ACHs from those for minihaloes, due to the increased rarity of
ACHs. The halo collapsed fraction is plotted in Figs
\ref{fig:meanfcoll-LMACH-114-64} and \ref{fig:meanfcoll-HMACH-114-64}. 
While LMACHs have a finite range in mass, because HMACHs are defined
to have a loose end, we assign their maximum mass as the one somewhat
smaller than the mass of the average-density cell: $M_{{\rm max}}=10^{11.5}\, M_{\odot}$
and $M_{{\rm max}}=10^{12.5}\, M_{\odot}$ for cells with $V_{{\rm cell}}=(1.78/h\,{\rm Mpc})^{3}$
and $V_{{\rm cell}}=(3.56/h\,{\rm Mpc})^{3}$, respectively. Otherwise,
the bias formalism breaks down (equation~\ref{eq:ps-bias}). Overall,
the mean values of both the LMACH collapsed fraction ($\left[f_{c,\,8:9}\right]_{a}$),
and the HMACH collapsed fraction ($\left[f_{c,\,9:11.5}\right]_{a}$
and $\left[f_{c,\,9:12.5}\right]_{a}$) are well predicted by equation~(\ref{eq:fcoll_bias})
when we adopt $\left(dn/dM\right)_{{\rm N-body},\, b}$ (equation
\ref{eq:Nbody-bias}). For LMACHs, $\left(dn/dM\right)_{{\rm ST},\, b}$
provides as good a fit as $\left(dn/dM\right)_{{\rm N-body},\, b}$,
except at $z=6$ where ST prescription somewhat overestimates the
mean. For HMACHs, the biggest discrepancy between $\left(dn/dM\right)_{{\rm ST},\, b}$
and $\left(dn/dM\right)_{{\rm N-body},\, b}$ exist at higher
redshifts (e.g. $z=15.6$) at $\log(1+\delta)\gtrsim0.$: here the
small number of sampled cells at high cell-density makes it difficult
to conclude which prescription provides a better estimator for the
mean bias. PS prescription provides a very poor fit at all redshifts.

The linear bias parameter, for both LMACHs and HMACHs, fails in predicting
the mean bias in general. This is noteworthy because even in the linear
regime, including the point $\delta=0$, the linear bias parameter
predicts the bias to be off from the observed values, which was also
the case for minihaloes. We discuss this issue in detail in Section
\ref{sub:Validity-of-Linear}.

In summary, even though LMACHs and HMACHs are very rare in the regime
we study, the nonlinear bias prescription combined with the mean N-body
halo mass function fits the observed mean halo bias very well throughout
the ranges of redshift and cell density we observe. Therefore, this
hybrid bias prescription can be applied for astrophysical and cosmological
applications in general.
We have indeed applied the bias prescription from this work in simulating
cosmic reionization by ACHs in a very large box, $425/h\,{\rm Mpc}$, in
order to populate Eulerian cells with size $425/h/504=0.843\,{\rm
Mpc}$ \citep{Iliev2014}. Because the halo mass resolution of the corresponding N-body simulation 
was only $10^{9} \,M_\odot$, we assigned each cell the missing LMACHs
using the mean conditional mass function
$\left(dn/dM\right)_{{\rm N-body},\,{\rm b}}$, where the LMACH
mean mass function from our $114/h\,{\rm Mpc}$ simulation was used to
generate $\left(dn/dM\right)_{{\rm N-body},\,{\rm b}}$.



%

%

\subsection{Stochasticity}

\label{sub:Result:scatter}

The average behaviour of conditional mass function is well understood
in terms of the biased mass function $\left(dn/dM\right)_{{\rm N-body},\,{\rm b}}$.
Now, how does the scatter of correlation around the mean compare to
the expected stochasticity? We showed in Section \ref{sub:Statistics}
that the variance of the number of haloes $N$ and given $\delta$
deviate from the simple Poisson value $\left[N\right]$ 
by the amount $\Delta_{{\rm scc}}(\delta)$. We now show the result
of simulation and compare this to the Poisson statistics
(equation~\ref{eq:concon_pdf_N}) and $\Delta_{\rm scc}(\delta)$
(equation~\ref{eq:scc}) by
explicitly calculating the sub-cell-scale
correlation function (equation \ref{eq:con_correl_definition}).

In Fig~\ref{fig:pdfMHcoarse} 
(see also Figs~5--9 in Supplementary Material)
we show 
the actual PDF and compare it to the expected Poisson distribution.
We find that the empirical PDF does not follow pure Poisson distribution
in general: the observed PDFs usually show large outliers compared
to the Poisson distribution, and there is no convincing case with
variance {\em smaller} than the Poissonian even though such a case
is possible if haloes are anti-correlated under given density environment
(equations~\ref{eq:scc}--\ref{eq:all_variance}). For example, Fig.~\ref{fig:pdfMHcoarse}
shows PDFs of minihalo population inside the $6.3/h$ Mpc box at different
redshifts and $\delta$'s. In order to get the distribution, each
chosen $\delta$ has some width $\Delta\delta$ such
that cells are chosen if their overdensity lies inside $\left[\delta-\Delta\delta/2,\,\delta+\Delta\delta/2\right]$.
$\Delta\delta$ is taken to be narrow enough to guarantee that the
PDF in each bin is a fair representation of the true PDF, while at
the same time wide enough to generate a large number of cells for
statistically reliable measure of the variance.

We also quantify the relative contribution of $\Delta_{{\rm scc}}$
to $\sigma^{2}(\delta)$ by the ratio $\Delta_{{\rm scc}}/\left[N\right]$
(note that it is compared not to $\left[N\right]_{a}$ but $\left[N\right]$),
in order to see the degree of deviation of PDF from the pure Poisson
distribution. There are several notable features in $\Delta_{{\rm scc}}/\left[N\right]$:
(1) At a given redshift, the ratio $\Delta_{{\rm scc}}/\left[N\right]$
decreases as mass of haloes increases, and thus LMACHs and HMACHs show
much weaker outliers progressively. Fig.~\ref{fig:pdfMHcoarse} 
and Figs~5--9 in Supplementary Material
show this trend: MHs have $\Delta_{{\rm scc}}/\left[N\right]\simeq[0,\,30]$,
LMACHs have $\Delta_{{\rm scc}}/\left[N\right]\simeq[0,\,6]$, and
HMACHs have $\Delta_{{\rm scc}}/\left[N\right]\simeq[0,\,2]$; (2)
As one increases the filtering scale -- or the size of cells -- $\Delta_{{\rm scc}}/\left[N\right]$
tends to decrease overall. We nevertheless have some exceptions in
this trend for MHs at very high-density cells ($\delta\simeq10$);
(3) $\Delta_{{\rm scc}}/\left[N\right]$ is not a monotonically increasing
or decreasing function of $\delta$; and (4) $\Delta_{{\rm scc}}$
is mostly positive both in underdense and overdense cells, indicating
that the sub-cell correlation is overall positive in both regimes
(see equation~\ref{eq:con_correl_definition}; this does not mean that
there is no negative values in $\overline{\xi_{12}}(\delta)$). This contradicts
the claim by \citet{Somerville2001}, where they usually find that
$\Delta_{{\rm scc}}<0$ in underdense regions and $\Delta_{{\rm scc}}>0$
in positive regions which led them to conclude that the correlation
function is negative inside underdense regions and positive in overdense
regions. We believe that this discrepancy comes from the erroneous
definition of $\Delta_{{\rm scc}}$ in \citealt{Somerville2001},
where they subtracted the global mean number of haloes $\left\langle N\right\rangle $
averaged over all cells of $\delta$ such that $\Delta_{{\rm scc}}=\sigma^{2}(\delta)-\left\langle N\right\rangle $,
while one should indeed define this as in equation~\ref{eq:all_variance}
to reflect the effect of the sub-cell correlation function.
The observed anti-correlation of $\overline{\xi_{12}}(\delta)$,
or negative values of $\overline{\xi_{12}}(\delta)$ when $r$ becomes comparable
to the cell size as seen in Fig.~\ref{fig:corr_MH45} 
(see also Figs~10--14 in Supplementary Material),
is due to  the finite cell size, because any correlation
existing inside a cell should be counter-balanced by anti-correlation
in order to conserve the halo number. 

We explicitly calculate $\overline{\xi_{12}}(\delta)$ defined by
equation (\ref{eq:con_correl_definition}) and $\Delta_{{\rm scc}}(\delta)$
from equation (\ref{eq:scc}). Toward this, we place a uniform grid with
$25^3$ sub-cells on each cell with $\delta$, such that
$dV_{1}=dV_{2}=V_{\rm cell}/25^3$. We then sample all sub-cell pairs with
given distance $r_{12}$ -- discretized as the distance between centres
of sub-cells -- and calculate $\overline{\xi_{12}}(\delta)$
using equation (\ref{eq:con_correl_definition}) and $\Delta_{{\rm
    scc}}(\delta)$ using equation (\ref{eq:scc})
We compare this value to the observed,
residual variance $\Delta_{{\rm scc}}=\sigma^{2}(\delta)-[N] $,
which are shown in the bottom panels of Fig.~\ref{fig:pdfMHcoarse},
denoted by $\Delta_{{\rm scc}\,(\int\xi)}$ and $\Delta_{{\rm scc\,(obs)}}$
respectively. The agreement between the two quantities are excellent,
and thus proves the fact that $\overline{\xi_{12}}(\delta)$ is the sole origin
for the super-Poissonian (or sometimes sub-Poissonian) variance in
$N(\delta)$  (see also Figs~5--9 in Supplementary Material). Due
to halo-number conservation, the correlation function 
is composed of positive (correlation) and negative (anti-correlation)
parts as seen in Fig.~\ref{fig:corr_MH45} (and Figs~10--14 in
Supplementary Material). 

\begin{figure*}
\includegraphics[width=160mm]{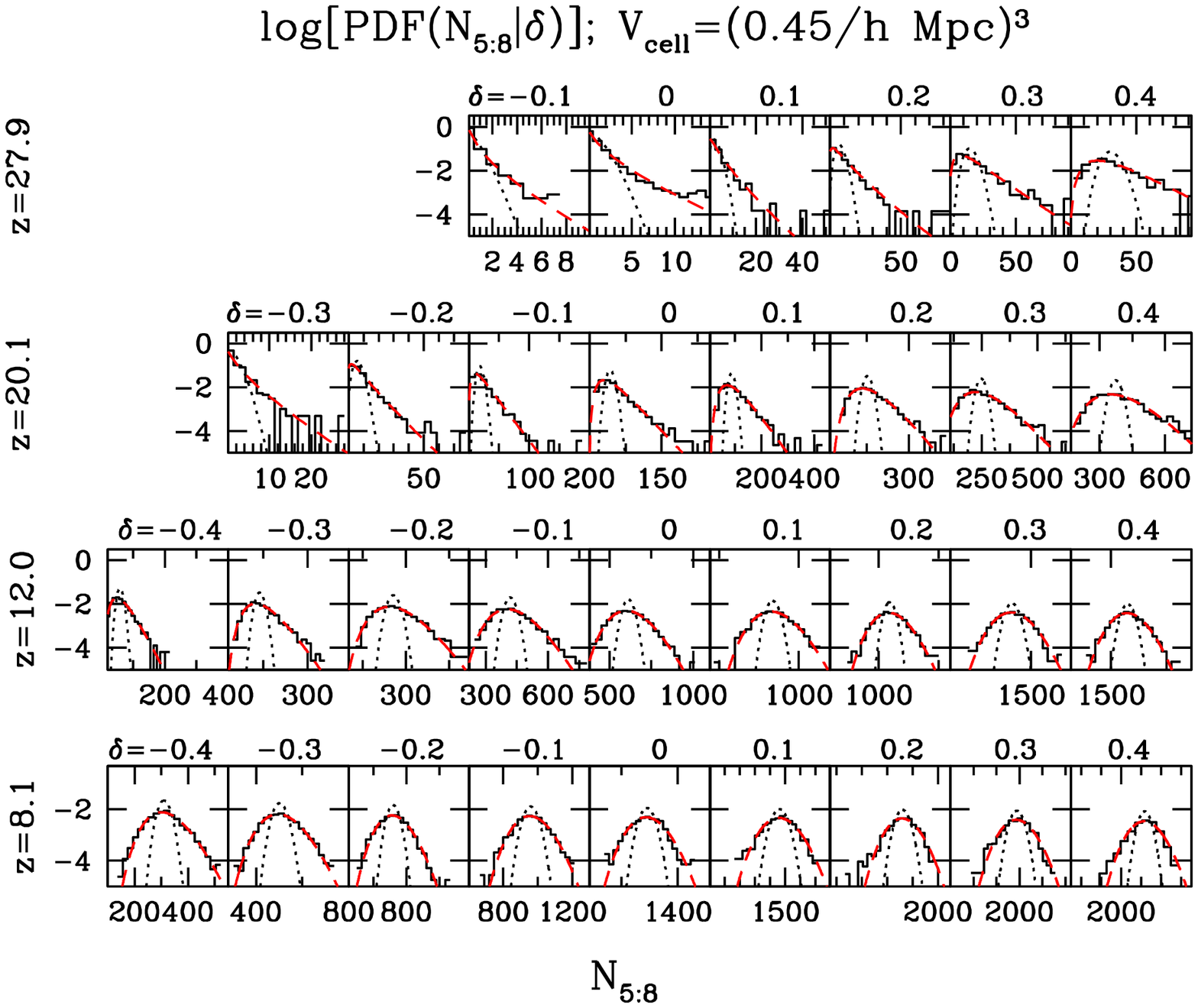}

\includegraphics[height=60mm]{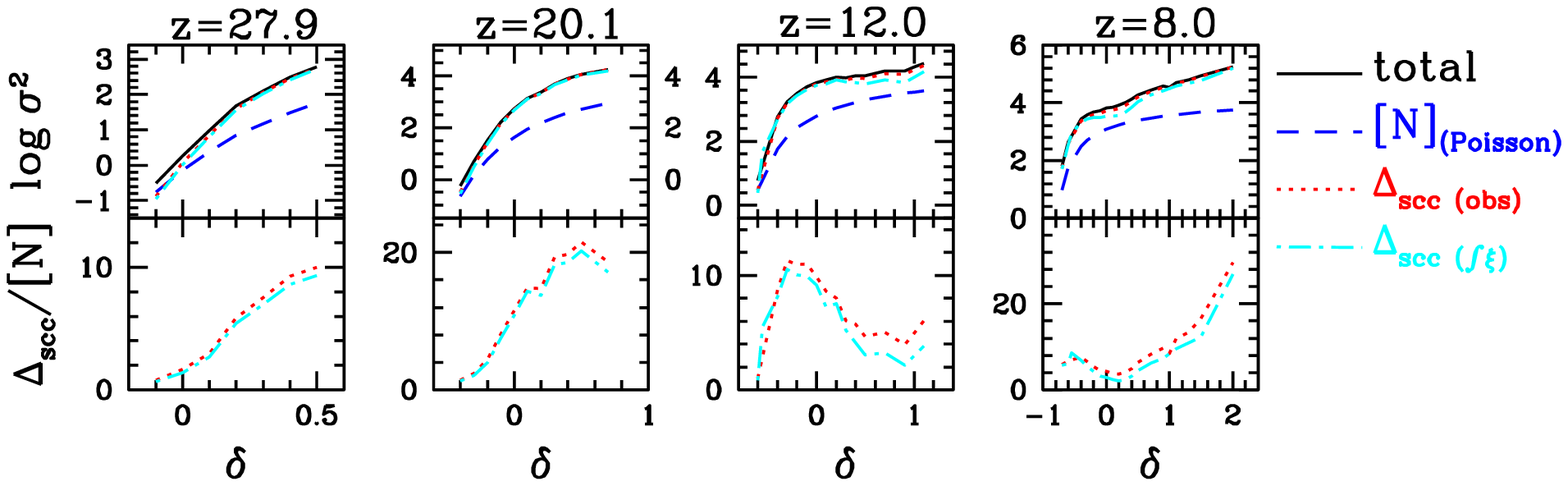}

\caption{(A) PDFs of the number of minihaloes $N_{5:8}$ at given overdensity
$\delta$ (denoted on top of each subpanel) of cells with Eulerian
volume $(0.45/h\,{\rm Mpc})^{3}$ observed in 20/h Mpc box. The horizontal
and vertical axes represent $N_{5:8}$ and $\log P_{{\rm
    cell}}(N_{5:8}|\delta)$, respectively.
The data from simulation (histogram) is compared to the pure
Poisson PDF (black, dotted) and the
super-Poissonian PDF (equation~\ref{eq:pdf_superPoi}; red,
dashed). (B) Variances. Plotted are
the observed total variance $(N-\left[N\right])^{2}$ (black, solid),
the purely Poissonian variance $[N]$ (blue, dashed),
the observed excess $\Delta_{{\rm
    scc\,(obs)}}=(N-\left[N\right])^{2}-\left[N\right]$ 
(red, dotted; eq. \ref{eq:con_correl_definition}), and
a value calculated from the sub-cell correlation function, $\Delta_{{\rm
    scc\,(\int\xi)}}$ (cyan, dot-dashed; eq. \ref{eq:scc}).
The range of $\delta$
and $z$ are selected such that the number of cells with given $\delta$
(binned properly as described in Section \ref{sub:Result:scatter})
at given $z$ exceeds 200 for the statistical reliability of the calculated
variances. The
ratio $\Delta_{{\rm scc}}/\left[N\right]$, plotted in the bottom
panels, quantifies the excess of variance over 
the purely Poissonian one, $\left[N\right]$. }
\label{fig:pdfMHcoarse}
\end{figure*}

\begin{figure*}
\includegraphics[height=140mm]{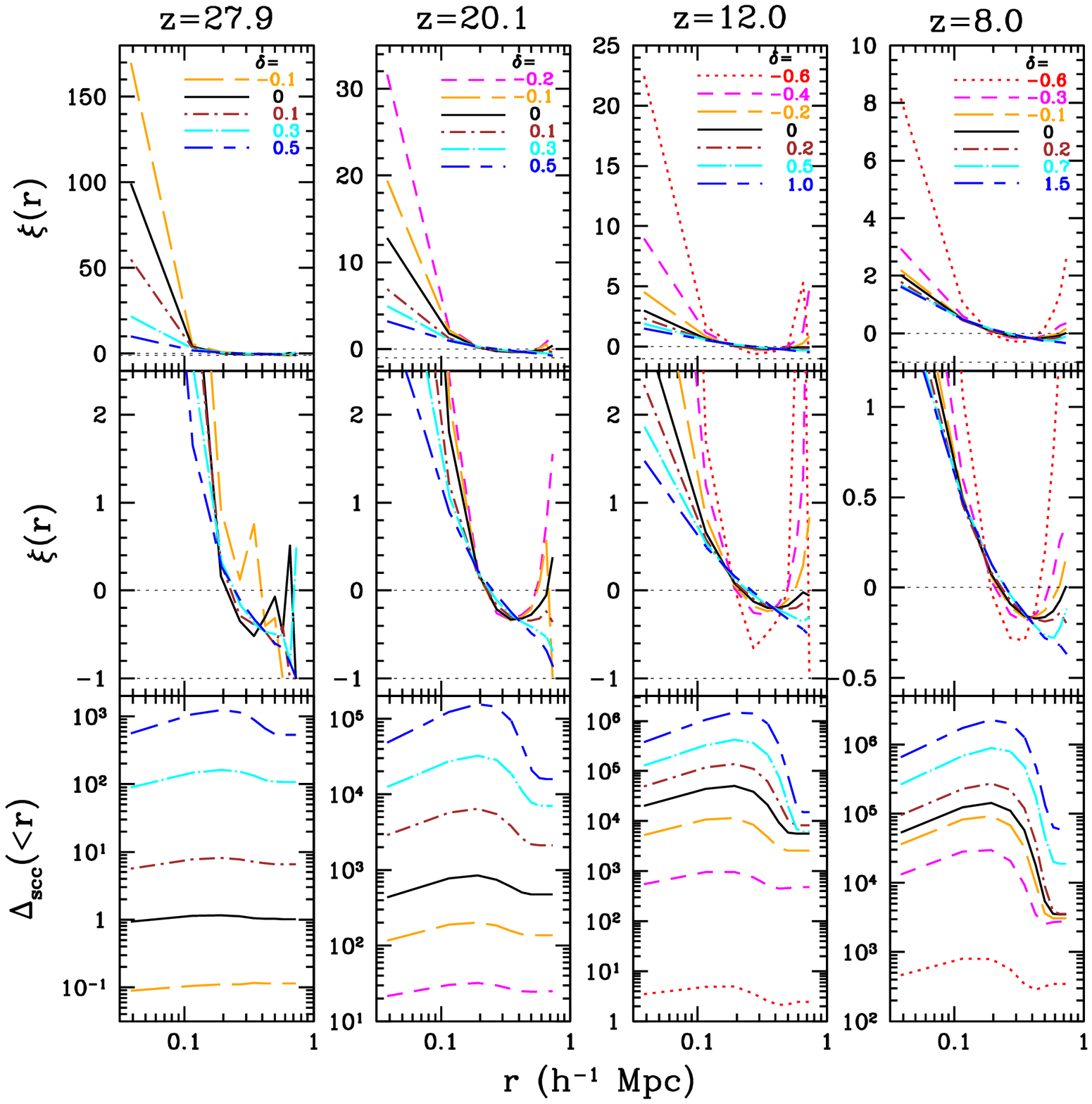}

\caption{Conditional sub-scale correlation function $\overline{\xi_{12}}(\delta)$
of minihaloes and the cumulative contribution to $\Delta_{{\rm scc}}(\delta)$,
inside $20/h\,{\rm Mpc}$ box with $V_{{\rm cell}}=(0.45/h)^{3}\,{\rm Mpc}$.
The
top- and middle-row subpanels show $\overline{\xi_{12}}(\delta)$ as a function
of two-point distance $r\equiv r_{12}$, and the bottom-row subpanels
show $\Delta_{{\rm
    scc}}(\delta;<r)\equiv\left(\frac{\left[N\right]}{V_{{\rm
      cell}}}\right)^{2}\int^{<r}dV_{1}dV_{2}\,\overline{\xi_{12}}(\delta;\,r_{12})$.}
\label{fig:corr_MH45}

\end{figure*}

\subsection{Bias in Perturbative Schemes}
\label{sub:Validity-of-Linear}

Local halo bias is often calculated or
fitted in perturbative way, i.e. as a polynomial series of
$\delta$:
\begin{equation}
\delta_{h}=\sum_{n=0}^{\infty}\frac{b^{(n)}}{n!}\delta^{n},
\label{eq:series}
\end{equation}
where the bias parameter $b^{(n)}$ is now defined as an $n$th-order moment in
this expansion. In practice, one should truncate the series by
limiting $\delta<1$ such that higher order moments decay more rapidly
than a few lowest-order moments. In this section, we re-visit the
perturbative scheme by MW and examine $b^{(n)}$ in more detail.

Linear bias approximation, $\delta_{h}\propto \delta$, is widely used
in literature and in practical 
applications such as galaxy surveys for cosmology. Here, the linear bias
parameter $b_{{\rm lin}}$ is useful when the mass of haloes is fixed,
because then $b_{{\rm lin}}$ is a simple constant coefficient for
varying $\delta$, or $\delta_{{\rm h}}=b_{{\rm lin}}\delta$, and
the same relation applies to $k$-space bias such that $\delta_{{\rm h}}({\bf k})=b_{{\rm lin}}\delta({\bf k})$.
Its limitation, however, has already been pointed out by MW themselves,
by expanding the nonlinear relation (equation \ref{eq:boost}) to
second order in $\delta$ and first order 
in $\sigma_{{\rm cell}}^{2}/\sigma_{M}^{2}$. 
Such expansion (and truncation at some order) is useful in observing
the halo bias in $k$-space because algebraic connection between real-space
parameters and $k$-space parameters is possible, and also in understanding
the generic behaviour of nonlinear bias. 
We therefore examine the Taylor-expanded
form of equation (\ref{eq:boost}). The main difference from MW is
that we expand the nonlinear relation to second order in $\delta$
but keeping $\sigma_{{\rm cell}}^{2}/\sigma_{M}^{2}$-dependence
accurate, because we sometimes reach 
$\sigma_{{\rm cell}}^{2}/\sigma_{M}^{2}\lesssim 1$.
This will enable us to examine the dependence of nonlinear bias on
the filtering scale more accurately.

We thus Taylor-expand $\delta_{{\rm h}}\left(M|\delta\right)$ to
the second order in $\delta$ while keeping the dependency on $R_{{\rm cell}}$
accurate (as in equation 30 of MW):
\begin{equation}
\delta_{{\rm h}}\left(M|\delta\right)=B_{0}+B_{1}\delta+\frac{1}{2}B_{2}\delta^{2},
\label{eq:2nd-expand}
\end{equation}
where we use $\delta_{{\rm lin}}=\delta+c\delta^{2}$ as an expansion
of $\delta_{{\rm lin}}$ ($c=-0.805$; see MW) and use the chain rule
$(\partial/\partial\delta)=(1+2c\delta)(\partial/\partial\delta_{{\rm lin}})$.
Using equations (\ref{eq:ps-bias}) and (\ref{eq:boost}), we obtain%
\footnote{Rigorously speaking, in this derivation, we assume that the filtering
scale $R_{{\rm cell},L}$ is fixed, and thus so is $\sigma_{{\rm cell}}^{2}$.
Because $(1+\delta)V_{{\rm cell}}=\frac{4 \pi}{3}R_{{\rm cell}}^{3}$, this means
that the $V_{{\rm cell}}$ changes as $V_{{\rm cell}}\propto(1+\delta)^{-1}$,
which is not compatible with the notion of uniform grid. If we were
to apply the expanded form on uniform-grid cases instead, $V_{{\rm cell}}$
is fixed and thus $R_{{\rm cell}}$ and $\sigma_{{\rm cell}}^{2}$
change as $\delta$ changes. Additional terms due to non-vanishing
$\left(\partial\sigma_{{\rm cell}}^{2}/\partial\delta\right)_{\delta=0}$
will appear on $B_{1}$ and $B_{2}$ in this case. Nevertheless, $\sigma_{{\rm cell}}^{2}$
is a very slowly-varying function in $\delta$ at $\left|\delta\right|\ll1$,
and thus we expect it to be higher-order correction in $\delta$,
and simply assume that 
$\left(\partial\sigma_{{\rm cell}}^{2}/\partial\delta\right)_{\delta=0}=0$
in the expansion in general.%
}
\begin{equation}
B_{0}=p^{-\frac{3}{2}}e^{-q}-1,\label{eq:B0}
\end{equation}
\begin{equation}
B_{1}=p^{-\frac{3}{2}}e^{-q}\left(1+\frac{p^{-1}\nu^{2}-1}{\delta_{c}}\right)
\label{eq:B1}
\end{equation}
and
\begin{equation}
B_{2}=p^{-\frac{3}{2}}e^{-q}\left\{
\frac{p^{-1}\nu^{2}}{\delta_{c}^{2}}\left(p^{-1}\nu^{2}-3\right)+\frac{2}{\delta_{c}}\left(p^{-1}\nu^{2}-1\right)\left(1+c\right)\right\}
,\label{eq:B2}
\end{equation}
where 
\begin{equation}
p\equiv1-\frac{\sigma_{{\rm cell},L}^{2}}{\sigma_{M,L}^{2}}
\label{eq:p_define}
\end{equation}
and 
\begin{equation}
q\equiv\frac{\nu^{2}}{2}\left(p^{-1}-1\right).
\label{eq:q_define}
\end{equation}
MW approximate the dependence on $\sigma_{{\rm cell},L}^{2}/\sigma_{M,L}^{2}$
to first order, and have $B_{0}=(\sigma_{{\rm cell},L}^{2}/2\sigma_{M,L}^{2})\left(3-\nu^{2}\right)$,
$B_{1}=b_{{\rm lin}}=1+(\nu^{2}-1)/\delta_{c}$, and $B_{2}=(\nu^{2}/\delta_{c}^{2})\left(\nu^{2}-3\right)+(2/\delta_{c})\left(\nu^{2}-1\right)\left(1+c\right)$,
to which equations (\ref{eq:B0}), (\ref{eq:B1}) and (\ref{eq:B2})
converge respectively when $\sigma_{{\rm cell},L}^{2}/\sigma_{M,L}^{2} \ll 1$.

$B_{0}$ explains the non-zero offsets $\left(dn/dM\right)_{b}(\delta=0)-\left\langle dn/dM\right\rangle $
and $f_{{\rm coll,b}}(\delta=0)-\left\langle f_{{\rm coll}}\right\rangle $
observed in almost all cases (see Figs \ref{fig:meannumber-mini-6.3-14}-\ref{fig:meanfcoll-HMACH-114-64}):
let us call this the ``0-point offset'' as MW did. 
0-point offset is a natural consequence of the fact that the
  global mean of a quantity $A$, $\left\langle A\right\rangle$,
differs from the selective average, $[A]_{\delta=0}$, only over cells
with $\delta=0$.
If one is to apply
a simple linear relation $\delta_{h}\propto \delta$,
it is presumed that $\left(dn/dM\right)_{b}(\delta=0)=\left\langle dn/dM\right\rangle $
(or $f_{{\rm coll,b}}(\delta=0)=\left\langle f_{{\rm coll}}\right\rangle $)
because $\delta_{{\rm h}}(\delta)=b_{{\rm lin}}\delta$ with $b_{{\rm lin}}$
as a constant coefficient. However, even in the linear regime in general,
$\delta_{{\rm h}}(\delta)=B_{0}+B_{1}\delta$ with non-zero $B_{0}$.
$B_{0}$ depends strongly on $\nu$. The negative sign of
$B_{0}$ reflects the fact that the rarer the haloes, or the 
higher the $\nu$, the smaller the chances are to find them in the
mean-density environment ($B_{0}<0$); in the opposite regime when
$\nu$ is small, $B_{0}>0$, which means that haloes are more abundant
in the mean-density cells than the mean value. The sign of $B_{0}$
also indicates, under a given filtering scale, the {}``overall''
tendency 
of halo distribution: when $B_{0}<0$, the net number of haloes found
in overdense regions is larger than that in underdense regions, and
when $B_{0}>0$, the net number of haloes found in overdense regions
is smaller than that in underdense regions. As a practical example,
it will be very important to study haloes in voids if those haloes are
a very abundant type, or $\nu\ll1$.

It is important to note that if the bias function (equation \ref{eq:mean_bias_def})
is expanded instead of $\delta_{{\rm h}}$, one should include the
singular term $B_{0}/\delta$ such that $b=B_{0}/\delta+B_{1}$,
because otherwise the approximated linear bias parameter cannot explain
the offset. In this sense, $b$ should not be taken as a physical
quantity but merely as a mathematical entity  representing the
fully nonlinear dependence of $\delta_{h}$ on $\delta$. $\delta_{{\rm h}}$
($=b\delta$) is a physical quantity which does not become singular
when $\delta\to0$.

\begin{figure}
\includegraphics[width=85mm]{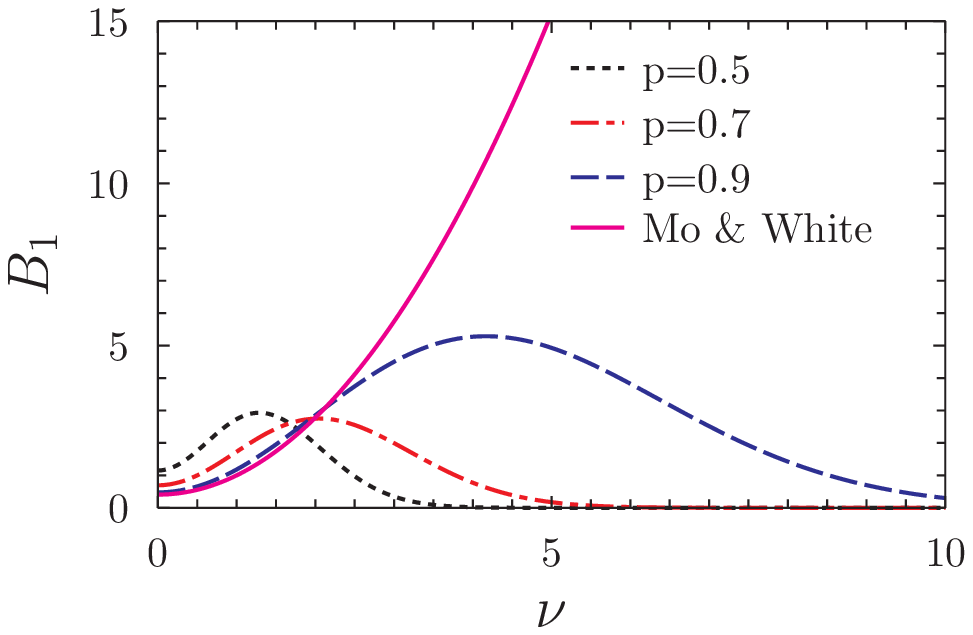}
\caption{Linear bias parameter $B_1$ as a function of $\nu$ and $p$
  (equation~\ref{eq:B1}). The linear bias parameter of MW
  corresponds to a case with $p=1$ ($q=0$ accordingly). Note that $B_1$ is in
  general a non-monotonic function of $\nu$: both
  very abundant ($\nu\ll 1$) and very rare ($\nu\gg 1$) haloes are
  weakly biased to the first order in $\delta$, while the commonly used $b_{\rm lin}=B_{1}(p=1)$ of
  MW is a monotonic function of $\nu$. Because $p\neq 1$ in
  practice, care needs to be taken when using $b_{\rm lin}$ for very
  rare haloes.}
\label{fig:B1-dependence}
\end{figure}

$B_{1}$ is a good indicator of the overall trend of bias. The sign
of $B_{1}$, which is always positive, guarantees that haloes are not
anti-biased but biased for higher cell-densities regardless of $\nu$
or the filtering scale, as long as $\delta$ is in the linear
regime. $B_{1}$ depends on both $\nu$ and 
$\sigma_{R_{{\rm cell,m}}}^{2}/\sigma_{M}^{2}$. 
At fixed halo-mass and filtering scale, $B_{1}$ increases as $\nu$
increases when $\nu<\nu_{{\rm crit,lin}}\equiv p\sqrt{1-\delta_{c}+2/(1-p)}$
and decreases when $\nu>\nu_{{\rm crit,lin}}$
(figure~\ref{fig:B1-dependence}). Such non-monotonic 
trend in $B_{1}$ would not be observed when filtering scale is large
enough, because then $\nu_{{\rm crit,lin}}\to\infty$. At fixed halo mass
(and thus fixed $\nu$ and $\sigma_{{\rm R}_{f}}^{2}$ at some $z$),
the effect of filtering scale or $p$ on $B_{1}$ is also a mixed
bag depending on rarity of haloes (or $\nu$) as seen in figure \ref{fig:B1-dependence}.

\section{Summary and Discussion}
\label{sec:conclusion}

We investigated the local bias of cosmological halo formation in the
fully nonlinear regime, using both halo data from N-body simulations
sampled on uniform grids and theoretical estimates for Eulerian halo
bias. Over the wide dynamic range of halo mass, from $10^{5}\,{\rm M}_{\odot}$
to $\sim10^{12}\,{\rm M}_{\odot}$, we find that the observed biased
population of haloes $\left(dn/dM\right)_{b}$ inside a cell
with density $\delta$ can be matched well by the convolution of
the mean N-body mass function $\left\langle dn/dM\right\rangle _{{\rm N-body}}$
with the nonlinear bias parameter derived from the extended Press-Schechter
formalism. Convolution with the PS mass function provides very
poor fits in general, and convolution with the ST mass function
provides fits slightly poorer than $\left\langle
dn/dM\right\rangle _{{\rm N-body}}$. Nevertheless, as the ST mass
function is known 
to break down for very rare haloes (see e.g. the large
discrepancy of the ST mass function for haloes of $M\ge
10^{6}\,M_\odot$ at $z\ge 20$ in Fig. \ref{fig:meanMF}), it is best to
avoid both PS and ST, and instead use $\left\langle
dn/dM\right\rangle _{{\rm N-body}}$ in convolving the mean mass
function to the bias factor given by equation (\ref{eq:boost}).
Based on the fact that the observed bias in halo population
is well matched by the hybrid estimate 
$\left(dn/dM\right)_{{\rm N-body},\, b}$ which combines two
physical quantities with different origins (the average mass function
$\left\langle dn/dM\right\rangle _{{\rm N-body}}$ is determined
by a specific halo-identification scheme and the nonlinear bias parameter
is based on the extended Press-Schechter theory), this prescription
should be applicable in general to cases under other
halo-identification schemes.

We also find that the variance of halo numbers inside grid cells with
given overdensity is not purely Poissonian, but has additional variance.
This variance originates from the sub-cell scale halo-halo correlation,
which we proved quantitatively by explicitly calculating the conditional
correlation functions. In the regime we studied ($z\gtrsim6$ and
uni-grid filtering with cell size of $\sim\left[0.2-3.6\right]\, h^{-1}\,{\rm Mpc}$),
we find that the additional variance is always positive except for
some negative values sporadically observed for haloes with
$M>10^{9}\,{\rm M_{\odot}}$.

The nonlinear bias prescription described in our paper can be used to
generate mock halo catalogues in the following sequence:

(i) Generate or adopt a mean mass function of haloes
$\left(dn/dM\right)_{\rm N-body}$. It is advised not to use
the PS mass function, due to the large discrepancy from the usual
N-body halo catalogues practically over the full mass range.

(ii) Generate a density field at a redshift of interest: if N-body
data is available, adopt a proper smoothing scheme to generate a
density field from the distribution of particles. Depending on
the size of cells, cell-density can become nonlinear, and therefore
N-body simulation is recommended.

(iii) Place a uniform grid on the density field from step (ii), and
identify the comoving volume of the cell as $V_{\rm cell}$.

(iv) Visit a cell, and identify the cell overdensity $\delta$. Use
equation (\ref{eq:mass-Rcell}) to deduce $R_{\rm cell}$. Take $R_{\rm cell}$ as
the spatial filtering scale of the linearly extrapolated density
field to $z=0$, and calculate the corresponding variance 
$\sigma^{2}_{R_{\rm cell}}$. Use equations (\ref{eq:lin-NL-pos}) and
(\ref{eq:lin-NL-neg}) (or the numerical fit given by 
equation 18 of MW) to find matching $\delta_{\rm lin}$ of
$\delta$. 

(v) To populate a cell with a halo of mass $M$, use equation
(\ref{eq:mass-R}) to obtain $R_f$, and take this as the filtering scale of
the lineaized density field at $z=0$ and calculate the corresponding
variance $\sigma^2_{M}$.

(vi) Plug quantities from steps (iv) and (v) in equation
(\ref{eq:ps-bias}), then use equation (\ref{eq:deltaH_lagrangian}), then
finally use equation (\ref{eq:Nbody-bias}) to calculate the biased
halo mass function 
$\left(dn/dM\right)_{{\rm N-body},\,b}$. 
Multiplying the infinitesimal mass bin $dM$ and $V_{\rm cell}$ 
to $\left(dn/dM\right)_{{\rm N-body},\,b}$, one obtains the
mean number of haloes $[N]$ of $M=[M,\,M+dM]$ in the cell.

(vii) Iterate steps (iv) - (vi) over all cells in the box.

(viii) If one wants to implement stochasticity, which should indeed
affect the power spectrum of halo density field, use equation
(\ref{eq:pdf_superPoi}) with $[N]$ from step (vi) and an empirically
found $\sigma^2(\delta)$ to include super-Poisson stochasticity and
sample halos by the Monte-Carlo method. For a selected 
range of halo masses and cell 
sizes as described in Sections~\ref{sub:Result:mean} and
\ref{sub:Result:scatter}, a reader may contact us for 
these values.

Perturbative approach to the nonlinear bias is found limited.
First, one needes to be careful when approximating the halo bias by a
simple linear relation $\delta_{h}\propto \delta$, because even when
the filtered density field is in the linear regime, $|\delta|\ll 1$, the 0-point
offset (equation~\ref{eq:B0}) may not be negligible. In such cases,
one should of course take $B_{0}$ into account such that $\delta\simeq
B_{0}+b_{\rm lin}\delta$.
This 0-point offset ($\left(dn/dM\right)_{b}(\delta=0)\neq\left\langle dn/dM\right\rangle $)
occurs in general when (1) haloes are rare and/or (2) the cell size
is small, which MW has already recognized and
we have confirmed from our data. In the nonlinear regime, even the
second-order perturbation, 
which we calculated without the approximation taken by MW
(Equations~\ref{eq:B0} -- \ref{eq:q_define}), provides
a very poor fit in general. We thus claim that the local nonlinear
bias scheme should be used unless perturbative approach is unavoidable.

Nonlinear bias schemes such as the one studied in this paper
can be applied to both cosmological and 
astrophysical problems. For example, we already used the mean bias
prescription in this paper as a sub-grid treatment to populate simulation
boxes with haloes which are not resolved otherwise, for simulating
cosmic reionization process: see \citet{Ahn2012} for populating
$114\, h^{-1}\,{\rm Mpc}$ box with minihaloes, and \citet{Iliev2014}
for populating $425\, h^{-1}\,{\rm Mpc}$ box with LMACHs. 
Similar approach has been attempted by \citet{delaTorre2013}, where
they test their bias-based sub-grid treatment against resolved N-body
haloes in terms of two-point statistics. 
Their bias prescription, however,
is heuristic and thus the corresponding fitting parameters should be
re-evaluated when e.g. a very different dynamic range of halo mass is
targeted.
In contrast, even though we have just studied cosmological
haloes at $z\gtrsim 6$, 
the agreement between data and theoretical prediction in such wide
range of halo mass, cell size, cell density, and redshift suggest
that this prescription is valid in general.

The nonlinear bias scheme studied here is valid when the primordial
density field is Gaussian, and thus may not be directly used to study
non-Gaussianity. It is also preferred that further study of the super-Poissonian
(or sometimes sub-Poissonian) stochasticity, which we quantified here
with $20\, h^{-1}\,{\rm Mpc}$ box for minihaloes and $114\, h^{-1}\,{\rm Mpc}$
box for LMACHs and HMACHs, is devised with higher-resolution, larger-box
simulations to increase statistical reliability. Stochasticity is
likely to have temporal correlation as well as spatial correlation,
which should be further studied for a more self-contained bias prescription.

\section*{Acknowledgments}

This work was supported by a research grant from Chosun University (2010).
All simulations in this work were undertaken at the Texas Advanced
Computing Center (TACC) at The University of Texas at Austin under
TeraGrid allocations.  


\end{document}